\documentclass[12pt,preprint]{aastex}

\newcommand{\myemail}{miurah@tap.scphys.kyoto-u.ac.jp}

\slugcomment{\today}

\shorttitle{Isotopic Fractionation during Shock-Waves Heating}
\shortauthors{Miura and Nakamoto}

\begin{document}


\title{Shock-Wave Heating Model for Chondrule Formation: \\
    Prevention of Isotopic Fractionation}


\author{Hitoshi Miura\altaffilmark{1,2} and Taishi Nakamoto\altaffilmark{3}}
\affil{Department of Physics, Kyoto University, Kitashirakawa, Sakyo, Kyoto 606-8502, Japan}
\email{\myemail}


\altaffiltext{1}{Department of Physics, Kyoto University}
\altaffiltext{2}{Research Fellow of the Japan Society for the Promotion of Science}
\altaffiltext{3}{Center for Computational Sciences, University of Tsukuba}


\begin{abstract}
Chondrules are considered to have much information on dust particles and processes in the solar nebula. It is naturally expected that protoplanetary disks observed in present star forming regions have similar dust particles and processes, so study of chondrule formation may provide us great information on the formation of the planetary systems. 

Evaporation during chondrule melting may have resulted in depletion of volatile elements in chondrules. However, no evidence for a large degree of heavy-isotope enrichment has been reported in chondrules. In order to meet this observed constraint, the rapid heating rate at temperatures below the silicate solidus is required to suppress the isotopic fractionation. 

We have developed a new shock-wave heating model taking into account the radiative transfer of the dust thermal continuum emission and the line emission of gas molecules and calculated the thermal history of chondrules. We have found that optically-thin shock waves for the thermal continuum emission from dust particles can meet the rapid heating constraint, because the dust thermal emission does not keep the dust particles high temperature for a long time in the pre-shock region and dust particles are abruptly heated by the gas drag heating in the post-shock region. We have also derived the upper limit of optical depth of the pre-shock region using the radiative diffusion approximation, above which the rapid heating constraint is not satisfied. It is about $1-10$. 

\end{abstract}





\section{INTRODUCTION}

Chondrules are millimeter-sized, once-molten, 
spherical-shaped grains mainly composed of silicate material. 
They are considered to have formed from chondrule precursor dust particles about $4.5\times10^9\,{\rm yr}$ ago in the solar nebula; 
they were heated and melted through flash heating events in the solar nebula 
and cooled again to solidify in a short period of time (e.g., Jones {\it et al}. 2000 and references therein). 
So they must have great information on the early history of our solar system. 
Since it is naturally expected that protoplanetary disks around young stars in star forming regions have similar dust particles and processes, the study of chondrule formation may provide us much information on the planetary system formation. 
Chondrules have many features: physical properties (sizes, shapes, densities, degree of magnetization, etc.), chemical properties (elemental abundances, degree of oxidation/reduction, etc.), isotopic compositions (oxygen, nitrogen, rare gases, etc.), mineralogical and petrologic features (structures, crystals, degrees of alteration, relicts, etc.), and so forth, each of which should be a clue that helps us to reveal their own formation process and that of the planetary system. Shock-wave heating model for chondrule formation has 
been studied 
by many authors (Hood \& Horanyi 1991, 1993, Ruzmaikina \& Ip 1994, Tanaka {\it et al.} 1998, Iida {\it et al}. 2001 (hereafter INSN), Desch \& Connolly 2002 (hereafter DC02), Miura {\it et al}. 2002, Ciesla \& Hood 2002 (hereafter CH02), Miura \& Nakamoto 2005 (hereafter MN05)), 
and is considered to be one of the plausible mechanisms for chondrule formation 
(Boss 1996, Jones {\it et al}. 2000). 

Tachibana \& Huss (2005) found that the degree of isotopic fractionation of sulfur in troilites (FeS) in chondrules from Bishunpur (LL3.1) and Semarkona (LL3.0) is quite small ($<0.1\%/{\rm amu}$ for all the grains, except one grain that has $0.27 \pm 0.14 \%/{\rm amu}$). The absence of isotopic fractionation in troilites suggests that chondrules have to be heated rapidly ($> 10^4 \, {\rm K/hr}$) at a temperature range of $1273 - 1473 \, {\rm K}$, in which the isotopic fractionation should occur associated with evaporation of troilites (Tachibana \& Huss 2005). Below $1273 \, {\rm K}$, troilites is solid state and it is assumed that no isotopic fractionation occurs when solid troilite evaporates because the time scale of sulfur diffusion in FeS is much larger than that of evaporation (the evaporation P\'{e}clet number for troilites at temperatures close to the eutectic point is about 100 for $50 \, {\rm \mu m}$-sized grain). On the contrary, above $1473 \, {\rm K}$, the melting troilite grains would have been surrounded by melted silicate. In this case, evaporation of sulfur would be controlled by diffusion through the surrounding silicate melt. However, since sulfur can hardly dissolve to the chondrule silicate melt, the troilite would not evaporate and a large degree of isotopic fractionation is not expected.  Dust aggregates that have never melted before are thought to be fluffy. When such aggregates are heated, the isotopic fractionation of troilites should occur associated with the evaporation from inside of fluffy aggregates. If the heating rate is slow, it is impossible to suppress the isotopic fractionation, because the duration of the evaporation becomes long enough to produce certain amount of isotopic fractionation. On the other hand, since once molten dust particles are not fluffy, they do not have to be heated rapidly to prevent from isotopic fractionation because FeS in the dust particle is completely covered by silicate components. Thus, we do not have to take into account the isotopic fractionation during the cooling phase. Namely, chondrules have to be heated rapidly at least in the first melting heating event. 

The isotopic fractionation can be suppressed due to not only the rapid heating but also the presence of back reaction from evaporated sulfur gas if the dust-to-gas mass ratio is large enough. However, in our situation, this effect can be negligible. Tachibana \& Huss (2005) also calculated the degree of isotopic fractionation of sulfur under conditions of sufficiently high dust-to-gas mass ratio in a closed system, and concluded that the required dust-to-gas mass ratio to suppress the isotopic fractionation by the back reaction is higher than about ten thousands times the solar value. It means that the dust-to-gas mass ratio should be greater than about 100. It has not been well investigated whether the shock-wave heating (the gas drag heating) works well in such extremely high dust-to-gas mass ratio environment or not, and to answer this problem is beyond the scope of this paper. Thus, we do not take into account the back reaction in this study. 

Tachibana {\it et al.} (2004) investigated the heating rate of chondrules in the framework of the shock-wave heating model and concluded that the gas drag heating in the post-shock region can heat chondrules rapidly enough to suppress the isotopic fractionation. However, it is known that chondrules are also heated in the pre-shock region due to the radiation emitted by ambient dust particles (DC02, CH02). The effect is well known as the blanket effect, which was not taken into account in the study by Tachibana {\it et al.} (2004). Results by DC02, in which the dust thermal radiation is taken into consideration as the radiation source, showed that the heating speed due to the radiation is too slow to suppress the isotopic fractionation ($\sim 300 \, {\rm K/hr}$). CH02 also performed numerical simulation taking the transfer of the dust thermal continuum emission into consideration. In contrast with DC02, results of CH02 showed that the heating rate of chondrules is large enough to suppress the isotopic fractionation even if the pre-shock region is dusty environment (dust-to-gas mass ratio is about 1.5, which is a few hundreds times larger than that of the solar abundance). 

In previous studies of shock-wave heating model taking into account the radiation transfer (DC02 and CH02), only the dust thermal continuum was taken into consideration as the radiation source. However, there is the other radiation source, the line emission of gas molecules. DC02 and CH02 neglected the line cooling because they assumed that the shock region is optically too thick to its own line emission to lose gas thermal energy effectively. On the contrary, MN05 showed an estimation that the post-shock gas is not so optically thick and the post-shock gas in a few $100-1000 \, {\rm km}$ behind the shock front can cool. Therefore, the line cooling should be taken into account. Moreover, numerical simulations in DC02 and CH02 have done for the limited shock conditions (the pre-shock gas number density is $n_0 \simeq$ a few $10^{14} \, {\rm cm^{-3}}$ and the shock velocity is $v_{\rm s} \simeq 7 - 9 \, {\rm km \, s^{-1}}$). These shock conditions are classified into the high-density and low-velocity shock waves in the appropriate shock condition for chondrule formation (INSN). Recently, a new shock wave generation mechanism which are sufficient to account for melting of dust particles was proposed (Nakamoto {\it et al.} 2005). In the case, X-ray flares generated at close to the protosun induce the shock waves in the upper solar nebula where the gas density is low. The typical shock condition of the new mechanism is estimated as $n_0 \sim 10^{11} \, {\rm cm^{-3}}$ and $v_{\rm s} \sim 40-60 \, {\rm km \, s^{-1}}$. There is no study for the low-density and high-velocity shock waves taking into account the thermal radiation from dust particles. 

The purpose of this paper is to develop a new shock-wave heating code taking into account the radiation transfer of both the line emission of gas molecules and the dust thermal radiation and to investigate how the thermal history of dust particles in the shock waves is affected by the optical properties of the pre-shock region. Especially, we focus the heating rate of precursor dust particles in a temperature range of $1273 - 1473 \, {\rm K}$, in which the isotopic fractionation should occur. The optical properties of the flow depend mainly on the dust size distribution and the dust-to-gas mass ratio. Moreover, we perform the simulation with various shock conditions (the pre-shock gas number density and the shock velocity) in order to discuss which shock generating mechanism is appropriate for chondrule formation. We describe details of our model and basic equations in section 2. We show the calculation results in section 3 and discussion in section 4. Finally, we summarize our study in section 5.

\section{MODEL AND BASIC EQUATIONS}


\subsection{Overview of Model}
\label{sec:overview_model}
Basic mechanism of the shock wave heating is rather simple. Let us suppose there is a gas medium containing dust particles with a dynamical equilibrium, i.e. they do not have a relative velocity initially. And let us suppose a shock wave passes the medium. Then, the gas is accelerated by the gas pressure and obtains some amount of velocity, while dust particles tend to remain the initial position. This causes the relative velocity between the dust particles and the gas. When the relative velocity is present, the frictional force and drag heating work on the dust particles; the intensity of the force and the heating depend mainly on the relative velocity and the gas density. Also, the high temperature gas in the post-shock region, heated by the compressional heating, heats the dust particles by thermal collisions. Moreover, dust particles are heated by radiation emitted from gas molecules and other dust particles. Dust particles are heated by those three processes, and cooled by emission of thermal radiation, the collision with cooler gas, and the latent heat cooling due to the evaporation. 

\begin{figure}
\epsscale{.90}
\plotone{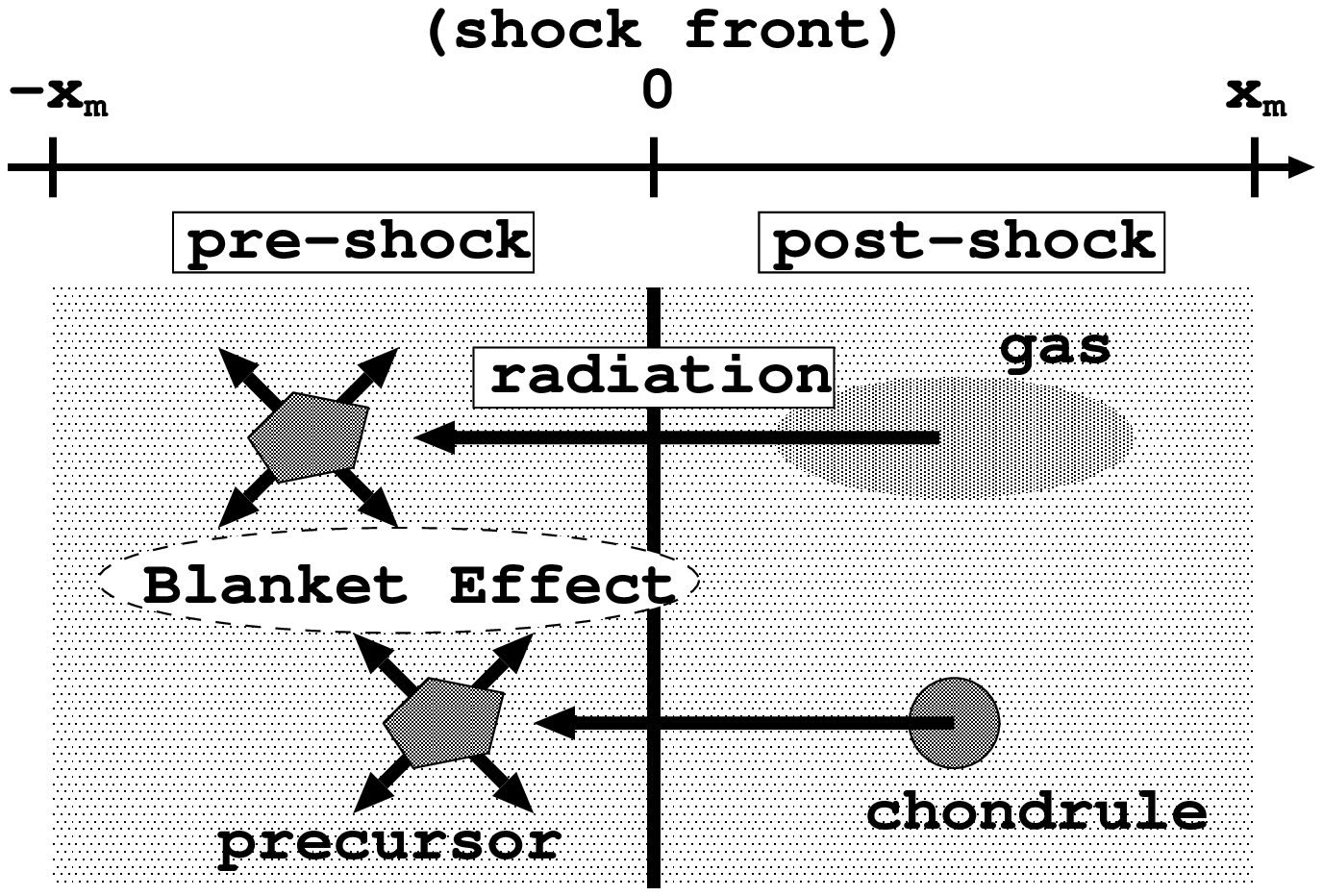}
\caption{A schematic view of our calculation region. Gas and precursor dust particles go together at the same speed from pre-shock boundary ($x = -x_{\rm m}$). They pass though the shock front ($x=0$), interact each other in the post-shock region, and finally reach the post-shock boundary ($x=x_{\rm m}$) and run away to the outer region in the nebula. The radiation emitted by the gas and dust particles in the post-shock region is absorbed by the dust particles in the pre-shock region. The dust particles in the pre-shock region also absorb the radiation emitted by other dust particles in the pre-shock region. In this paper, we call this effect the blanket effect. \label{fig:blanket_effect}}
\end{figure}

\begin{figure}
\epsscale{.80}
\plotone{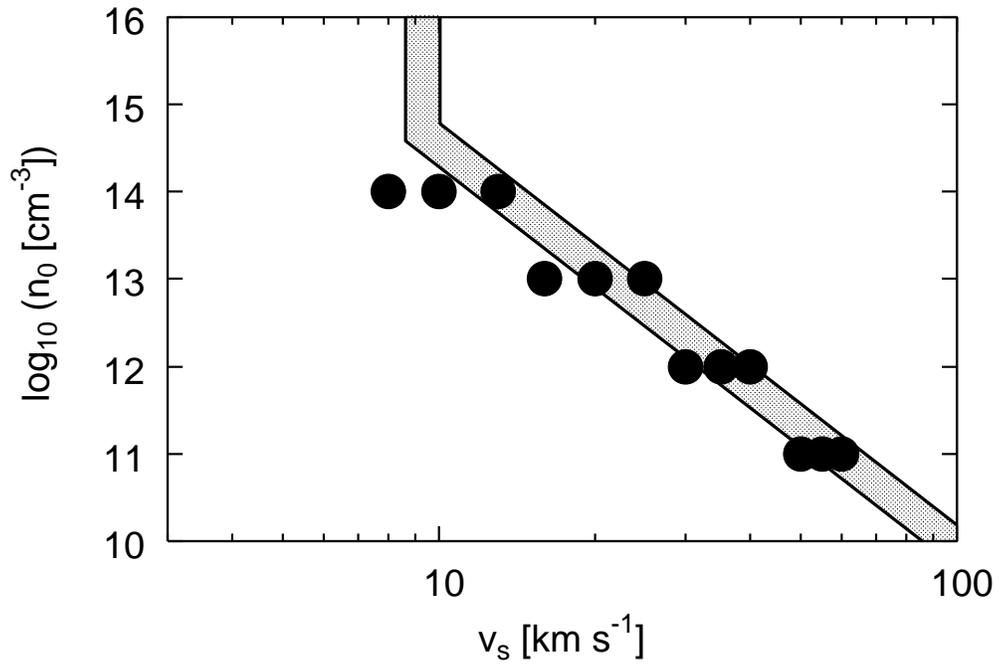}
\caption{Shock conditions for which we perform simulations are displayed as a function of the shock velocity $v_{\rm s}$ and the pre-shock gas number density $n_0$. Gray region indicates the shock condition which is sufficient to heat precursor dust particles up to $1473 - 2000 \, {\rm K}$ (INSN). \label{fig:shock_condition}}
\end{figure}

We have developed a new numerical model that simulates 1-dimensional plane-parallel steady shock flow including dust particles. The new model is based on our previous work (MN05) and added the thermal radiation transfer part to the model of MN05. We use the spatial coordinate, which is perpendicular to the shock front, in the shock front rest frame (see Fig. \ref{fig:blanket_effect}). The spatial extent of the calculation region is assumed to be the same as DC02 for comparison ($x_{\rm m} = 10^5 \, {\rm km}$). As will be shown later (sec. \ref{sec:extended_shock}), the choice of $x_{\rm m}$ is not a fundamental problem in this study. The origin is situated at the shock front. Initially, gas and dust particles flow together at the same speed (shock velocity) from $x = -x_{\rm m}$. In the pre-shock region, dust particles are gradually heated by the radiation as they approach the shock front. We focus the heating rate of the dust particles in this phase. In order to calculate the radiation intensity in the pre-shock region, we also have to evaluate the post-shock structures because the radiation mainly comes from the gas and dust particles in that region. When the gas flow arrives at the shock front, it is suddenly compressed and decelerated by the post-shock gas pressure. We consider the J-type shock and the jump condition is given by the Rankine-Hugoniot relation. The temperature and the density of the gas in the post-shock region are determined by the atomic and molecular cooling processes, chemical heating/cooling, and energy transfer between the gas and the dust particles. After passing through the shock front, dust particles have a relative velocity to the gas flow. The gas drag heating due to the relative velocity increases the dust temperature. In order to form chondrules, dust particles must be heated as high as the melting temperature. Appropriate shock condition for chondrule formation, in which precursor dust particles are heated enough to melt and do not vaporize completely, is numerically and analytically derived by INSN as a function of the shock velocity, $v_{\rm s}$, and the pre-shock gas number density, $n_0$. In this study, we simulate with twelve different shock conditions in which chondrules can be formed (see Fig. \ref{fig:shock_condition}). 

Since line emission of gas molecules also heats dust particles, we have to estimate how much energy of line emission comes from the post-shock region. It is expected that some fraction of the line emission emitted from the post-shock region is absorbed by gas molecules before escaping toward the pre-shock region. Previous studies of shock-wave heating are classified into two extreme cases; one is that the line cooling is perfectly neglected ({\it e.g.}, DC02 and CH02) and the other is that a constant fraction of line emission escapes from the post-shock region regardless of the column density of gas molecules ({\it e.g.,} INSN). DC02 suggested that the shock region is optically too thick to the line emission to lose more than $\sim 10\%$ of its energy. On the other hand, INSN considered that since the dust-melting region is very near from the shock front, the line cooling takes place at the region because the optical depth between the emitting region and the shock front is not so large. The real situation should be somewhere between these two models. In this study, we calculate the cooling rate due to the line emission taking the column density of gas molecules into consideration. Since a half of emitted photons goes toward upstream and the rest photons go toward downstream, the column density for each direction is different. Details are described in Appendix \ref{sec:net_cooling_rate}. 

The gas cooling due to the line emission contributes to a part of the source term of the radiation field. DC02 and CH02 solved the radiation transfer of the dust thermal continuum emission, however, they did not take into account the line emission of gas molecules. We add the source term due to the line emission to the radiation model. We consider that only the photons which can finally escape from the post-shock region contribute the radiation field. This is equivalent to the on-the-spot approximation (Osterbrock 1989) used in radiation transfer and photoionization calculations for H II regions. Details of formulation is described in Appendix \ref{sec:emission_coefficient}. The radiation affects the thermal histories of dust particles, and simultaneously depends on it. Therefore, we have to calculate the gas/dust dynamics and the radiation transfer consistently. The procedure is iterated until the mean intensity does not change more than 1\%.

\subsection{Gas Dynamics\label{sec:gas_dynamics}}
The heating and cooling processes violate the adiabatic condition of the gas. Then, 1-dimensional, steady flow is governed by the conservation equations of mass and momentum and the energy equation as follows: 
\begin{equation}
\rho_0 v_{\rm s} = \rho v ,
\label{eq:mass_conservation}
\end{equation}
\begin{equation}
\rho_0 v_{\rm s}^2 + p_0 = \rho v^2 + p ,
\label{eq:momentum_conservation}
\end{equation}
\begin{equation}
\frac{de}{dt} = \frac{e+p}{\rho} \frac{d\rho}{dt} - \Lambda, 
\label{eq:energy_gas}
\end{equation}
where $v_{\rm s}$ is the shock velocity, $\rho$, $v$, $p$, and $e$ are the density, velocity, pressure of the gas and the gas internal energy per unit volume, respectively. The subscript ``0" stands for values at the upstream boundary of the calculation ($x=-x_{\rm m}$). The net cooling rate $\Lambda$ contains the atomic/molecular cooling processes, chemical heating/cooling processes, and energy transfer between the gas and dust particles, and is given by
\begin{equation}
\Lambda = \Lambda_{\rm Ly\alpha} + \Lambda_{\rm H_2O(V)} + \Lambda_{\rm CO(V)} + \Lambda_{\rm H_2O(R)} + \Lambda_{\rm CO(R)} + \Lambda_{\rm dust} + \Lambda_{\rm H_2diss} - \Gamma_{\rm H_2form}, 
\end{equation}
where Ly$\alpha$, H$_2$O(V), CO(V), H$_2$O(R), and CO(R) stand for the cooling processes due to the line emission of Lyman $\alpha$, H$_2$O vibration, CO vibration, H$_2$O rotation, and CO rotation, respectively. The energy transfer rate from gas to dust particles is $\Lambda_{\rm dust}$. The cooling (heating) processes associated with the H$_2$ dissociation (formation) is $\Lambda_{\rm H_2diss}$ ($\Gamma_{\rm H_2form}$). The system of rate equations which governs nonequilibrium chemical reactions among gas species is written as 
\begin{equation}
\frac{dy_i}{dt} = n_{\rm H_0} \sum_{j=1}^{28} \sum_{k=1}^{28} k_{jk} y_j y_k + n_{\rm H_0}^2 \sum_{l=1}^{28} \sum_{m=1}^{28} \sum_{n=1}^{28} k_{lmn} y_l y_m y_n ,
\label{eq:chemical_reaction}
\end{equation}
where $n_{\rm H_0}$ is the total number density of hydrogen nuclei, $y_i$ is the relative abundance of species $i$ to $n_{\rm H_0}$, $k_{jk}$ and $k_{lmn}$ are the reaction rate coefficients, respectively. The first and second terms in Eq.(\ref{eq:chemical_reaction}) describe the two-body and three-body interactions in the gas phase, respectively. We include 156 reactions among 28 gas species. The interaction of particles with radiation is not included. Included processes in Eq.(\ref{eq:chemical_reaction}) and adopted rate coefficients are listed in INSN\footnote{INSN included 176 reactions among 35 gas species. In this study, we neglect D- and Si-inclusive species because they do not contribute to the gas cooling processes as coolant species. }. The time derivatives in Eq.(\ref{eq:energy_gas}) and (\ref{eq:chemical_reaction}) are Lagrangian derivatives. We numerically integrate Eqs.(\ref{eq:mass_conservation})-(\ref{eq:energy_gas}) and (\ref{eq:chemical_reaction}) using a finite difference algorithm described in Shapiro and Kang (1987). 

When the gas flow meets the shock front, it is suddenly compressed and decelerated by the pressure of gas in the post-shock region. The jump condition is given by the Rankine-Hugoniot relation, 
\begin{equation}
\frac{\rho_2}{\rho_1} = \frac{v_1}{v_2} = \frac{(\gamma +1) {\cal M}^2}{(\gamma -1) {\cal M}^2 + 2}, 
\label{eq:rankinehugoniot01}
\end{equation}
\begin{equation}
\frac{T_2}{T_1} = \frac{ \{2 \gamma {\cal M}^2 - (\gamma -1) \} \{ (\gamma -1) {\cal M}^2 +2 \}}{(\gamma +1)^2 {\cal M}^2},
\label{eq:rankinehugoniot02}
\end{equation}
where $T$ is the temperature of gas and $\gamma$ is the ratio of specific heat of gas. The subscript ``1" and ``2" stand for immediately before and behind the shock front, respectively. The Mach number in the pre-shock region is expressed as ${\cal M} \equiv v_1/\sqrt{\gamma k_{\rm B} T_1 / \bar{m}}$, where $\bar{m}$ is the mean molecular weight of the gas and $k_{\rm B}$ is the Boltzmann constant. 

In Eqs. (\ref{eq:mass_conservation}) and (\ref{eq:momentum_conservation}), we neglect the mass loading by evaporation of dust materials and the momentum exchange with dust particles. This assumption is valid because we consider low dust-to-gas mass ratio ($\le 0.1$, see sec. \ref{initial_condition}). However, we take into account the energy exchange with dust particles in Eq. (\ref{eq:energy_gas}) because the dust cooling for the gas can work well even in the low dust-to-gas mass ratio environment.

\subsection{Dust Dynamics}
If there is a relative velocity of dust particles to the gas, dust motion is changed by the gas drag force. 
The equation of motion is written by 
\begin{equation}
\frac{4}{3} \pi a_{\rm d}^3 \rho_{\rm d} \frac{dv_{\rm d}}{dt} = - \pi a_{\rm d}^2 \frac{C_{\rm D}}{2} \rho | v_{\rm rel}| v_{\rm rel},
\end{equation}
where $\rho_{\rm d}$, $v_{\rm d}$, $C_{\rm D}$, and $v_{\rm rel}$ are the dust particle material density, the velocity of dust particle, the drag coefficient, and the relative velocity of dust particle to the gas, respectively. The expression of drag coefficient is given in some previous papers ({\it e.g.}, Probstein 1968, Hood \& Horanyi 1991, INSN). 

The dust temperature changes due to some heating/cooling mechanisms; energy transfer with the gas, absorption of ambient radiation, radiative energy loss, and latent heat cooling due to the evaporation. The energy equation for the dust particle which governs the dust temperature is written as 
\begin{equation}
\frac{4}{3} \pi a_{\rm d}^3 \rho_{\rm d} C \frac{dT_{\rm d}}{dt} = 4 \pi a_{\rm d}^2 (\Gamma_{\rm g-p} + \Gamma_{\rm rad} - \Lambda_{\rm rad} - \Lambda_{\rm evap}), 
\end{equation}
where $C$ and $T_{\rm d}$ are the specific heat of dust particles and the dust temperature, respectively. Regarding the heating rate of the energy transfer with gas $\Gamma_{\rm g-p}$, we adopt the same expression of previous studies (Probstein 1968, Hood \& Horanyi 1991). The heating/cooling rates due to the absorption of ambient radiation $\Gamma_{\rm rad}$, the emission of dust thermal continuum radiation $\Lambda_{\rm rad}$, and the latent heat cooling by evaporation $\Lambda_{\rm evap}$ are given as
\begin{equation}
\Gamma_{\rm rad} = \epsilon_{\rm abs} \pi {\cal J}, \quad \Lambda_{\rm rad} = \epsilon_{\rm emit} \sigma_{\rm SB} T_{\rm d}^4, \quad \Lambda_{\rm evap} = J_{\rm evap} L_{\rm evap}, 
\end{equation}
where $\epsilon_{\rm abs}$ ($\epsilon_{\rm emit}$), ${\cal J}$, $\sigma_{\rm SB}$, $J_{\rm evap}$, and $L_{\rm evap}$ are the Planck-mean absorption (emission) coefficient, the mean intensity of ambient radiation, the Stefan-Boltzmann constant, the evaporation rate of dust component per unit area, and the latent heat of dust component, respectively. We assume $\epsilon_{\rm abs} = \epsilon_{\rm emit}$ and use the optical properties of astronomical silicate (Miyake \& Nakagawa 1993). If the dust temperature increases as high as the boiling temperature, the boiling takes place in the molten droplet. After reaching the boiling point, the dust temperature does not change because the input energy is consumed for the phase transition from liquid to gas. Assuming bubbles generated in molten dust particle go away to outside quickly, the boiling leads to shrinkage of the dust particle (Miura {\it et al.} 2002). 

The particle radius decreases due to the evaporation from the dust surface. 
The evolution of the particle radius due to evaporation is given by (Miura {\it et al.} 2002)
\begin{equation}
- \frac{da_{\rm d}}{dt} = \frac{J_{\rm evap}}{\rho_{\rm d}}. 
\end{equation}
We use the evaporation rate of forsterite given by Tsuchiyama {\it et al.} (1998). 

Finally, we should mention the fragmentation of molten dust particle due to the ram pressure. 
In the post-shock region, the ram pressure acts on the surface of molten dust particles. 
Susa \& Nakamoto (2002) estimated the maximum radius of molten dust particles above which the ram pressure dominates the surface tension and the molten dust particle should be fragmented. 
In our study, if the dust radius is larger than the maximum radius during its melting phase, we assume that the dust particle is divided into two molten droplets which have the same volume.

\subsection{Mean Intensity of Radiation Field}
We are considering the steady, one-dimensional, plane-parallel structure for the chondrule-forming region. The radiation transfer equation along the ray inclining with the angle $\theta$ against the spatial coordinate $x$ is written as (see Eq. \ref{eq:radtr04})
\begin{equation}
\mu \frac{d{\cal I}}{dx} = -(\alpha_{\rm p} + \sigma_{\rm p}) ( {\cal I} + {\cal S} ), 
\label{eq:radtr}
\end{equation}
where $\mu$ is equal to $\cos \theta$, ${\cal I}$ is the frequency-initegrated specific intensity of the radiation field, $\alpha_{\rm p}$ ($\sigma_{\rm p}$) is the Planck-mean absorption (scattering) coefficients of dust particles, and ${\cal S}$ is the frequency-integrated source function. The absorption/scattering coefficients are given by integrating the absorption/scattering cross sections with dust radius as 
\begin{equation}
{ \alpha_{\rm p} \choose \sigma_{\rm p} } = \int n_{\rm d} \pi a_{\rm d}^2 { \epsilon_{\rm abs} \choose \epsilon_{\rm scat} } da_{\rm d}, 
\end{equation}
where $n_{\rm d}$ is the number density of dust particles per unit radius per unit volume and $\epsilon_{\rm scat}$ is the planck-mean scattering efficiency. The source function includes three terms; dust thermal continuum emission, the line emission of gas molecules, and the scattering (see Eq. \ref{eq:source_function}). If the scattering is taken into consideration, the radiation intensity ${\cal I}$, which is the solution of the radiation transfer equation, is included in the source function. Therefore, we calculate the radiation transfer equation iteratively until the solution converges. 

The source term of the dust thermal continuum emission $j_{\rm d}$ (see Eq. \ref{eq:source_function}) is given as the total contribution of all dust sizes; 
\begin{equation}
j_{\rm d} = \int n_{\rm d} a_{\rm d}^2 \epsilon_{\rm emit} \sigma_{\rm SB} T_{\rm d}^4 da_{\rm d} .
\end{equation}
DC02 assumed that sub-micron sized dust particles are dynamically well coupled to the gas and are always in thermal equilibrium with the gas and the radiation field, however, we solve the dust thermal histories of various sized dust particles ($a_{\rm d} = 0.01 \, {\rm \mu m} - 1 \, {\rm cm}$) without the assumption of the thermal equilibrium. 

The source term of the line emission $j_{\rm g}'$ (see Eq. \ref{eq:source_function}) is given as the summation of all cooling mechanisms due to the line emissions. The line emission would be absorbed or scattered by the gas molecules while the photon is traveling from the emitting point to a certain point at which we evaluate know the radiation intensity. As a result, the value of $j_{\rm g}'$ decreases as the distance between two points increases more and more. In our model, the absorption of the line emissions by gas molecules is considered based on Neufeld \& Kaufman (1993) and the effect is taken into account in the term of $j_{\rm g}'$ (see Appendix \ref{sec:emission_coefficient}). 

Here, we define the optical depth as 
\begin{equation}
\tau (x) = \int_{-x_{\rm m}}^x ( \alpha_{\rm p} + \sigma_{\rm p} )dx.
\label{eq:optical_depth}
\end{equation}
Given the incident radiation fields and the source function at all the places ${\cal S}$, the mean intensity of radiation ${\cal J}$ can be found (Mihalas \& Mihalas 1999): 
\begin{equation}
{\cal J} (\tau) = \frac{{\cal I}_{\rm pre}}{2} E_2 (\tau) + \frac{1}{2} \int_0^{\tau_{\rm m}} {\cal S} (\tau') E_1 ( | \tau-\tau' | ) d\tau' + \frac{{\cal I}_{\rm post}}{2} E_2 (\tau_{\rm m} - \tau) ,
\label{eq:mean_intensity}
\end{equation}
where $E_n$ is the exponential integrals; $E_n (x) \equiv \int_1^\infty y^{-n} e^{-xy} dy$, ${\cal I}_{\rm pre}$ and ${\cal I}_{\rm post}$ are given as the boundary conditions of radiation intensity at $x = - x_{\rm m}$ and $x = x_{\rm m}$, respectively. We set ${\cal I}_{\rm pre} = {\cal I}_{\rm post} = \sigma T^4 / \pi = 1.46 \times 10^5 \, {\rm erg \, cm^{-2} \, ster^{-1} \, s^{-1}}$, which corresponds to the black body radiation with temperature of $300 \, {\rm K}$. Notice that we calculate the value of $j_{\rm g}$ for any set of $\tau$ and $\tau'$ based on Neufeld \& Kaufman (1993). We can also solve for the net flux of radiation energy ${\cal F}$ (Miharas \& Mihalas 1999): 
\begin{eqnarray}
{\cal F} (\tau) &=& 2 \pi {\cal I}_{\rm pre} E_3 (\tau) + 2 \pi \int_0^\tau {\cal S} (\tau') E_2 (\tau-\tau') d\tau' \nonumber \\
 &-& 2 \pi \int_\tau^{\tau_{\rm m}} {\cal S} (\tau') E_2 (\tau'-\tau) d\tau' - 2 \pi {\cal I}_{\rm post} E_3 (\tau_{\rm m} - \tau) .
\label{eq:radiative_flux}
\end{eqnarray}

\subsection{Physical Parameters}
The initial gas temperature is assumed to be 300 K. The initial gas-phase elemental abundance (ratio by number to hydrogen nuclei) is taken from Finocchi et al. (1997) for the solar nebula: $y_{\rm H} = 10^{-5}$, $y_{\rm He} = 9.75 \times 10^{-2}$, $y_{\rm CO} = 1.065 \times 10^{-4}$, and $y_{\rm H_2O} = 5.978 \times 10^{-4}$. Other hydrogen nuclei are assumed to exist as hydrogen molecules. Other species = 0. Moreover, it is assumed that the heavy elements like Si are in the dust particles and do not exist in the gas phase. 

The precursor dust particles are assumed to be composed of only forsterite with the mass density $\rho_{\rm mat} = 3.22 \, {\rm g \, cm^{-3}}$ (Saxene et al. 1993). The specific heat $C = 1.42 \times 10^7 \, {\rm erg \, g^{-1} \, K^{-1}}$, the latent heat of evaporation $L_{\rm evap} = 1.12 \times 10^{11} \, {\rm erg \, g^{-1}}$, and the latent heat of boiling $L_{\rm boil} = 1.64 \times 10^{11} \, {\rm erg \, g^{-1}}$ (NIST WebBook are adopted).\footnote{NIST WebBook. A gateway to the data collections of the National Institute of Standards and Technology. Available at http://webbook.nist.gov.} For the values of absorption/emission/scattering coefficients ($\epsilon_{\rm abs}$/$\epsilon_{\rm emit}$/$\epsilon_{\rm scat}$), we use the optical properties of astronomical silicate (Miyake \& Nakagawa 1993).

\subsection{Initial Conditions for Calculations\label{initial_condition}}
Appropriate shock conditions for chondrule formation, in which the precursor dust particles are heated enough to melt and do not vaporize completely, is numerically and analytically derived by INSN as a function of the pre-shock gas number density $n_0$ and the shock velocity $v_{\rm s}$. Fig. \ref{fig:shock_condition} shows the appropriate shock condition which is sufficient to heat the precursor dust particles up to $1473 - 2000 \, {\rm K}$ (gray region). Based on their results, we select some set of the $n_0$ and $v_{\rm s}$ for the initial conditions: $v_{\rm s} =$ 8, 10, 13 km s$^{-1}$ for $n_0 = 10^{14} \, {\rm cm^{-3}}$, $v_{\rm s} =$ 16, 20, 25 km s$^{-1}$ for $n_0 = 10^{13} \, {\rm cm^{-3}}$, $v_{\rm s} =$ 30, 35, 40 km s$^{-1}$ for $n_0 = 10^{12} \, {\rm cm^{-3}}$, and $v_{\rm s} =$ 50, 55, 60 km s$^{-1}$ for $n_0 = 10^{11} \, {\rm cm^{-3}}$ (filled circles). We abbreviate the shock condition of $n_0 = 10^{14} \, {\rm cm^{-3}}$ and $v_{\rm s} = 10 \, {\rm km \, s^{-1}}$ as ``n14v10", and so forth (see Table \ref{table:shock_condition}). 

\begin{figure}
\epsscale{.90}
\plotone{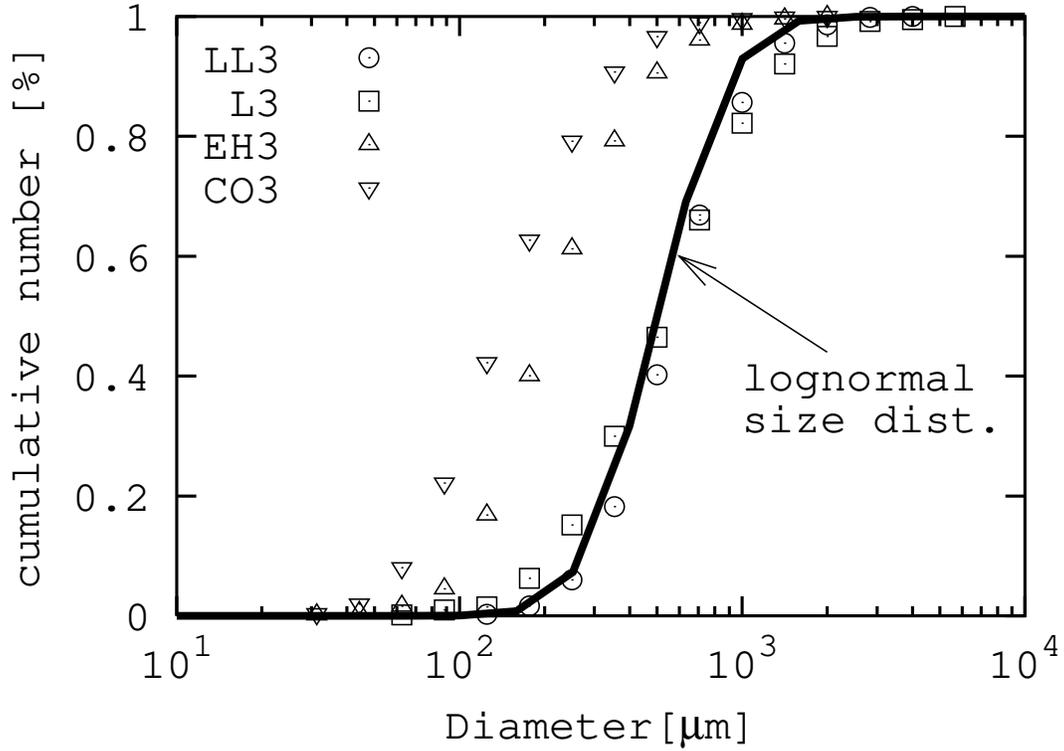}
\caption{Initial dust size distribution model we adopted in this study is plotted in case of lognormal one. The vertical axis is normalized cumulative number of chondrules whose radii are smaller than that of horizontal axis. The thick solid curve is the lognormal size distribution that we adopt in this study. Measured chondrule size distributions are also displayed for comparison; each symbol means results of LL3 chondrites by Nelson \& Rubin (2002) (LL3), L3 chondrites by Rubin \& Keil (1984) (L3), EH3 chondrites by Rubin \& Grossman (1987) (EH3), and CO3 chondrites by Rubin (1989) (CO3), respectively. \label{fig:size_dist}}
\end{figure}

The radiation field could be significantly affected by the blanket effect due to dust particles. Thus, the optical properties of the flow, which are mainly determined by the dust-to-gas mass ratio and the initial size distribution of precursor dust particles, should be paid attention in our study. We assume two size distribution models; power-law distribution and lognormal one. In the power-law one, the dust number density per unit radius per unit volume is given as $n_{\rm d} (a_{\rm d}) \propto a_{\rm d}^{-m}$, in which we assume $m=3.5$ in this study. The lognormal one is similar to the size distribution of chondrules in ordinary chondrites (Fig. \ref{fig:size_dist}). We assume that the size range of dust particles is from $0.01 {\rm \mu m}$ to $1 \, {\rm cm}$. We divide it into 31 bins and calculate time evolution of those particles simultaneously with gas dynamics. We change the dust-to-gas mass ratio in the pre-shock region as 0.01, 0.03, and 0.1. Totally, we have six cases as the dust particle model for each shock condition. The dust models are summarized in Table \ref{table:dust_model}.




\section{RESULTS}
We first show the effect of line emission on the gas dynamics in sec. \ref{sec:line_cooling}. Second, how the radiation field and dust thermal histories are affected by the optical depth is shown in sec. \ref{sec:blanket_effect} and \ref{sec:dust_thermal_history}. Finally, we focus on the heating rate of dust particle in the pre-shock region in sec. \ref{sec:heating_rate}.

\subsection{Line Cooling \label{sec:line_cooling}}
Figure \ref{fig:heatcoolrate_gas} shows the abundance of hydrogen molecule (top), the gas temperature (middle), and the heating/cooling rates for gas (bottom) in the post-shock region. The horizontal axis is the distance from the shock front. The shock condition is n14v10 in Fig. \ref{fig:shock_condition} and the dust model is L01 in Table \ref{table:dust_model}. After passing through the shock front, the gas temperature jumps up to $4420 \, {\rm K}$. Then the dissociation of hydrogen molecule takes place and the gas cools rapidly. At a few tens km behind the shock front, the dissociation of hydrogen molecule ceases and the formation dominates the dissociation instead. Then 
the heating due to the formation of hydrogen molecule balances with the cooling due to the dissociation of it, so the temperature plateau is formed at $T_{\rm g} \simeq 2000 \, {\rm K}$, on which the gas temperature does not change significantly. In this phase, we find that the line cooling plays an important role; the line cooling due to the H$_2$O vibrational emission can remove the H$_2$ formation energy effectively. Thus, the abundance of hydrogen molecule increases gradually during the gas temperature is on the plateau. If no line cooling exists, the cooling mechanism which can remove the formation energy is only the dust cooling. However, its cooling rate is too low to cool the gas effectively. The formation energy released until all the hydrogen nuclei become hydrogen molecules is estimated as $E = y_{\rm H_2} n_{\rm H_0} E_{\rm form} \sim 10^3 \, {\rm erg \, cm^{-3}}$, where $y_{\rm H_2} \sim 0.1$, $n_{\rm H_0} \sim 10^{15} \, {\rm cm^{-3}}$, and $E_{\rm form} = 7.17 \times 10^{-12} \, {\rm erg}$ is the formation energy per molecule, and the cooling rate by dust particles is about $\Lambda_{\rm dust} \simeq10^{-2} \, {\rm erg \, cm^{-3} \, s^{-1}}$ (see the bottom panel in Fig. \ref{fig:heatcoolrate_gas}), so it takes about $10^5 \, {\rm s}$ to remove the formation energy. Since the gas velocity is about $1 \, {\rm km \, s^{-1}}$ in the post-shock region, the gas would run away from the calculation region before their energy is removed completely. Therefore, the line emission is important for the gas cooling. Moreover, this result indicates that the line emission is an important radiation source.

\subsection{Blanket Effect \label{sec:blanket_effect}}
We think that it is helpful for readers to show the structure of radiation field in the chondrule-forming region. Because of the blanket effect, the radiation intensity strongly depends on the optical depth, which is specified by the size distribution and the dust-to-gas mass ratio $C_{\rm d}$. Fig. \ref{fig:flux} shows the mean intensity ${\cal J}$ (top) and the radiation flux ${\cal F}$ (bottom) as a function of the distance from the shock front. The shock model is n14v10 and the dust models are L10 ($C_{\rm d} = 0.10$), L03 ($C_{\rm d} = 0.03$), and L01 ($C_{\rm d} = 0.01$). We find that the more the pre-shock region becomes optically thick, the more strongly the radiation intensity is amplified by the blanket effect. This result is naturally expected. The mean intensity immediately in front of the shock front is ${\cal J} = 6.36$ for L10, $3.68$ for L03, and $2.85$ for L01 in unit of $10^7 \, {\rm erg \, cm^{-2} \, ster^{-1} \, s^{-1}}$, respectively. The temperature of dust particles which are in the thermal equilibrium with the radiation field is given as $T_{\rm rad} = ( \pi {\cal J} / \sigma_{\rm SB} )^{1/4}$. Hereafter, we call it the radiation temperature. For each case, the radiation temperature is $T_{\rm rad} = 1370 \, {\rm K}$ for L10, $1195 \, {\rm K}$ for L03, and $1121 \, {\rm K}$ for L01, respectively. Therefore, in the case of L10, the dust temperature in the pre-shock region exceeds $1273 \, {\rm K}$, which is the melting point of FeS, and the isotopic fractionation begins to take place. Since the heating rate due to the blanket effect is lower than $10^3 \, {\rm K/hr}$, the isotopic fractionation will be observed if chondrules have been formed in such dusty shock waves. On the contrary, in cases of L03 and L01, the radiation temperature does not exceed $1273 \, {\rm K}$ in the pre-shock region, so the isotopic fractionation begins to take place after passing though the shock front. Behind the shock front, the gas drag heats dust particles very rapidly ($>10^5 \, {\rm K/hr}$) up to $1473 \, {\rm K}$, above which the isotopic fractionation is expected not to occur because the silicate melt surrounds FeS melt and suppresses its evaporation. 

The radiation flux is also an important information to understand our study. In the bottom panel of Fig. \ref{fig:flux}, it is found that the radiation flux is almost constant in the pre-shock region. In the case of L10, the pre-shock region is optically thick ($\tau_{\rm pre} = 2.45$), where $\tau_{\rm pre}$ is defined as the optical depth of the pre-shock region; 
\begin{equation}
\tau_{\rm pre} = \int_{-x_{\rm m}}^0 ( \alpha_{\rm p} + \sigma_{\rm p} ) dx. 
\end{equation}
The reason why the radiation flux in the pre-shock region is almost constant in spite of the optically thick environment is the re-emission of dust thermal continuum emission (another explanation about the constancy of the radiation flux is shown in Appendix \ref{appen:constancy_radiation_flux}). Since the dust temperature near the shock front is higher than that far from the shock front, the re-emitted dust thermal continuum emission produces net radiation flux toward upstream. This effect can be understood with the radiative diffusion approximation (Rybicki \& Lightman 1979). We discuss it more in sec. \ref{sec:radiative_diffusion}.

\subsection{Dust Temperature in Pre-Shock Region\label{sec:dust_thermal_history}}
Figure \ref{fig:dust_temp} shows dust thermal histories in the pre-shock region. Calculation conditions are the same as those of Fig. \ref{fig:flux}; the shock condition is assumed to be n14v10, and dust models are L10, L03, and L01. A horizontal dotted line indicates $1273 \, {\rm K}$, above which the isotopic fractionation takes place until the dust temperature exceeds $1473 \, {\rm K}$. In cases of L03 and L01, dust temperatures do not exceed $1273 \, {\rm K}$ in the pre-shock region, so the isotopic fractionation does not take place in this region. After passing through the shock front, dust particles are rapidly heated by the gas drag heating. It takes about $1 \, {\rm s}$ for dust temperature to increase from $1273 \, {\rm K}$ to $1473 \, {\rm K}$, so the heating rates become about $10^6 \, {\rm K/hr}$ in both cases. It is enough large to suppress the isotopic fractionation. On the contrary, in the case of L10, the dust temperature exceeds $1273 \, {\rm K}$ in the pre-shock region at $x \simeq - 2 \times 10^4 \, {\rm km}$. This is due to the ambient radiation field which is amplified by the blanket effect. After that, it takes about $2 \times 10^3 \, {\rm s}$ for dust particles to meet the shock front. After passing through the shock front, dust temperature increases up to $1473 \, {\rm K}$ within $1 \, {\rm s}$, so the duration is negligible compared to $2 \times 10^3 \, {\rm s}$. Therefore, the heating rate between $1273 \, {\rm K}$ and $1473 \, {\rm K}$ is estimated as about $360 \, {\rm K/hr}$. It is too slow to suppress the isotopic fractionation, so it should take place in the case of L10. 

Here, we compare our results with that of DC02 and CH02, which took into account the radiation transfer of dust thermal continuum emission, but not the gas line emission (line cooling). The canonical case of DC02 had following parameters; the initial gas density is $10^{-9} \, {\rm g \, cm^{-3}}$, the shock velocity is $7 \, {\rm km \, s^{-1}}$, and the dust-to-gas mass ratio is 0.005. Despite the low dust-to-gas mass ratio, the pre-shock dust temperature reaches about $1700 \, {\rm K}$ at the shock front (see Fig. 4 of DC02). We guess that it is because of their treatment of the sub-$\mu$m sized dust particles in the source function (Eq. 5 of DC02). They used the gas temperature in the source term of sub-$\mu$m sized dust particles. However, dust temperature is not perfectly the same as the gas temperature due to the ambient radiation field and the radiative cooling of dust particle themselves. For example, in the case of Fig. \ref{fig:heatcoolrate_gas}, the post-shock gas temperature is $2049 \, {\rm K}$ at $x \simeq 10^3 \, {\rm km}$, where the $\mu$m-sized dust particles are in thermal equilibrium with the gas and the radiation field because there is no relative velocity between the gas and $\mu$m-sized dust particles. The $\mu$m-sized dust temperature at this point is $1247 \, {\rm K}$ according to our results. The radiation source term is proportional to the forth power of the temperature. Thus, DC02 overestimated the source term of sub-$\mu$m sized dust particles about by one order of magnitude. On the contrary, we think that the model of CH02 underestimated the radiation intensity in the pre-shock region because they did not take into account the line emission of the gas. As we showed in sec. \ref{sec:line_cooling}, some amount of the line emission emitted in the post-shock region can penetrate into the pre-shock region. Considering the line emission, the radiation intensity in the pre-shock region should be larger than that of the model without this effect.

\subsection{Heating Rate \label{sec:heating_rate}}
We calculated the thermal histories of dust particles for various shock conditions and dust models and obtained the dust temperature immediately in front of the shock front and the heating rate of dust particles in a temperature range of $1273 - 1473 \, {\rm K}$, in which the isotopic fractionation occurs. Our results are summarized in Tables \ref{table:summary_a100}$-$\ref{table:summary_a398}. The heating rate, $R_{\rm heat}$, is defined as 
\begin{equation}
R_{\rm heat} \equiv \frac{1473 \, {\rm K} - 1273 \, {\rm K}}{\Delta t}, 
\end{equation}
where $\Delta t$ is the duration in which dust temperature increases from $1273 \, {\rm K}$ to $1473 \, {\rm K}$ in the heating phase. In Fig. \ref{fig:Tpre}(a), we plot the results of the dust temperature just in front of the shock front $T_{\rm sf}$ as a function of the optical depth of the pre-shock region. In shock conditions n14v10 and n13v20, it is found that the more the pre-shock region is optically thick, the higher the dust temperature just in front of the shock front becomes. If the optical depth of the pre-shock region $\tau_{\rm pre}$ exceeds about 2, the pre-shock dust temperature exceeds $1273 \, {\rm K}$, so the isotopic fractionation takes place in the pre-shock region. On the contrary, in shock conditions n12v35 and n11v55, the dust temperatures just in front of the shock front seem not to depend on $\tau_{\rm pre}$ because the optical depth is very small for any dust models. The maximum value of $\tau_{\rm pre}$ is 0.14 for n12v35 and 0.013 for n11v55. In such an optically thin environment, the blanket effect does not work well. Fig. \ref{fig:Tpre}(b) shows the heating rates of dust particles $R_{\rm heat}$. It is found that if the pre-shock dust temperature exceeds $1273 \, {\rm K}$, the heating rate decreases drastically. In these cases, the heating rate is about a few $100 \, {\rm K/hr}$ and it is too slow to suppress the isotopic fractionation. On the contrary, if the pre-shock dust temperature does not exceed $1273 \, {\rm K}$, the heating rate is very large ($\sim 10^6 \, {\rm K/hr}$). In these cases, the heating rate is determined by the gas drag heating in the post-shock region and the isotopic fractionation can be suppressed by the rapid heating. From these results, we find that the heating rates are clearly separated into two cases; the very slow heating ($R_{\rm heat} < 10^3 \, {\rm K/hr}$) and the very rapid heating ($R_{\rm heat} \sim 10^6 \, {\rm K/hr}$). The critical optical depth of the pre-shock region above which the isotopic fractionation should occur is between 1.56 and 2.45 in the shock condition of n14v10.

\section{DISCUSSIONS}
In sec. \ref{sec:heating_rate}, we found that the more the pre-shock region is optically thick, the higher the pre-shock dust temperature becomes. Moreover, we found that the heating rates of dust particles tend to drop drastically when the dust temperatures just in front of the shock front, defined as $T_{\rm sf}$, exceed $1273 \, {\rm K}$. Although this condition might not be satisfied any time, we think that $T_{\rm sf}$ can be a good indicator in order to judge whether the rapid heating constraint is satisfied or not\footnote{One of the exceptions is thought to be the shock waves with very small spatial scale, such as the planetesimal bow shock. We discuss that in sec. \ref{sec:planetesima_bow_shock}.}. 

According to our simulation results, the critical optical depth of the pre-shock region above which $T_{\rm sf}$ exceeds $1273 \, {\rm K}$ is between 1.56 and 2.45 for shock condition of n14v10. In sec. \ref{sec:radiative_diffusion}, we derive the pre-shock dust temperature analytically using the radiative diffusion approximation. In sec. \ref{sec:critical_optical_depth}, we estimate the upper limit of the optical depth using discussion in sec. \ref{sec:radiative_diffusion}. We discuss cases of larger $x_{\rm m}$ value assuming spatially extended shock waves (sec. \ref{sec:extended_shock}). We mention the cooling rate constraint for chondrule formation in sec. \ref{sec:cooling_rate} and which shock wave generating mechanism is appropriate for chondrule formation in sec. \ref{sec:shock_mechanism}. We also consider the effect of porous structures of precursor dust particles on the dust thermal histories in sec. \ref{sec:porous}. Finally, we discuss the validity of the one-dimensional plane-parallel geometry in sec. \ref{sec:validity_plane-parallel} and the rapid heating constraint in sec. \ref{sec:validity_rapid_heating}.

\subsection{Pre-shock Dust Temperature \label{sec:radiative_diffusion}}
In the bottom panel of Fig. \ref{fig:flux}, it is found that the radiation flux is almost constant in the pre-shock region. In the case of L10, the pre-shock region is optically thick ($\tau_{\rm pre} = 2.45$). In this case, it is expected that the radiation flux emitted from the post-shock region decreases by a factor of about $e^{-2.45} \sim 0.086$ until it reaches the pre-shock boundary ($x=-x_{\rm m}$). The reason why the radiation flux is almost constant in spite of the optically thick environment is the re-emission of dust thermal continuum emission. The dust temperature near the shock front is higher than that far from the shock front, so the re-emitted dust thermal continuum emission produces net radiation flux toward upstream. In the optically thick environment, the equation of radiative diffusion (Rosseland approximation) gives the relation between the radiative flux and the temperature gradient as (Rybicki \& Lightman 1979)
\begin{equation}
{\cal F} = - \frac{4 \sigma_{\rm SB}}{3} \frac{d T_{\rm d}^4}{d \tau}, 
\label{eq:radiative_diffusion}
\end{equation}
where $d \tau = \kappa_{\rm R} dx$ and $\kappa_{\rm R}$ is the Rosseland mean absorption coefficient. In this discussion, we assume $\kappa_{\rm R} = \kappa$, for simplicity. Assuming ${\cal F}$ to be constant and integrating Eq. (\ref{eq:radiative_diffusion}) from $x=-x_{\rm m}$ to $x=0$, we obtain the dust temperature just in front of the shock front $T_{\rm sf}$ as 
\begin{equation}
T_{\rm sf}^4 = T_0^4 + \frac{3}{4\sigma_{\rm SB}} |{\cal F}| \tau_{\rm pre}, 
\label{eq:temperature_shockfront}
\end{equation}
where $T_0$ is the dust temperature at $x = -x_{\rm m}$ and estimated by assuming the thermal equilibrium with the radiation field as $T_0 = (\pi {\cal J} / \sigma_{\rm SB})^{1/4}$. At $x=-x_{\rm m}$, since the radiation intensity coming from outer region ($x<-x_{\rm m}$) is weaker than that from the chondrule-forming region, we neglect its effect. Assuming the isotropic intensity in an angle range of $\pi/2 < \theta < \pi$, the mean intensity at $x = -x_{\rm m}$ is given as ${\cal J} = {\cal I}/2 = |{\cal F}|/(2\pi)$. Next, the radiation flux ${\cal F}$ can be estimated by a following consideration. The origin of the radiative energy is the gas kinetic energy entering into the shock front. The gas kinetic energy flux is approximately estimated as $\frac{1}{2} \rho_0 v_{\rm s}^3$. Some fraction of the gas energy flux returns toward upstream as a form of radiation flux. Assuming the fraction is $f$, we obtain $|{\cal F}| = \frac{f}{2} \rho_0 v_{\rm s}^3$. Our calculation results of $f$ are summarized in Tables \ref{table:summary_a100}-\ref{table:summary_a398}. Finally, we can estimate the dust temperature immediately in front of the shock front as  
\begin{equation}
T_{\rm sf} = \bigg( \frac{2 + 3\tau_{\rm pre}}{4 \sigma_{\rm SB}} \frac{f}{2} \rho_0 v_{\rm s}^3 \bigg)^{1/4}. 
\label{eq:temperature_shockfront3}
\end{equation}
In Fig. \ref{fig:Tpre2}, we compare the pre-shock dust temperature estimated by Eq. (\ref{eq:temperature_shockfront3}) with the calculated results for various shock conditions: n14v10 (top left), n13v20 (top right), n12v35 (bottom left), and v11v55 (bottom right), respectively. We substitute $\rho_0 = \bar{m} n_0$ into Eq. (\ref{eq:temperature_shockfront3}), where the mean molecular weight $\bar{m} = 2.35 \times 10^{-24} \, {\rm g}$ resulting from the initial gas abundance we adopted in this study and draw three curves which correspond to $f=0.3$, $0.5$, and $1.0$, respectively. Triangles indicate the calculation results for the log-normal dust size distribution and circles are for power-law one. The calculation results seem to well match with the analytic formula (Eq. \ref{eq:temperature_shockfront3}) in cases of $\tau_{\rm pre} \ga 1$ if the difference of $f$ among each dust model is taken into consideration. On the contrary, in the optically thin environment ($\tau_{\rm pre} \la 1$), especially in the shock conditions of n12v35 and n11v55, $T_{\rm sf}$ of calculation results are larger than a curve indicating $f=1$ which means that all of the gas energy flux flowing into the shock front is converted into the radiation energy flux and it returns upstream. The reason why the calculation results exceed the upper limit is thought to be due to the optically thin environment in which the radiative diffusion approximation cannot be applied. However, our results show that the analytic formula Eq. (\ref{eq:temperature_shockfront3}) is valid for the relatively optically thick environment ($\tau_{\rm pre} \ga 1$) in which the radiative diffusion approximation can be used. 

We would like to mention the effect of the radiation intensity coming from outer region ($x < -x_{\rm m}$), which is neglected in above estimation. In an optically thick environment, the radiation intensity is proportional to $T_{\rm d}^4$. We assume the radiation intensity coming from outer region corresponds to the black body radiation of $300 \, {\rm K}$. If the typical dust temperature in chondrule-forming region is $1200 \, {\rm K}$, the ratio of the radiation intensity coming from outer region to that in the chondrule-forming region becomes only about 0.004. Therefore, our assumption that the radiation intensity coming from outer region is negligible is valid. Even if the temperature in outer region is $500 \, {\rm K}$, its contribution is only about 0.03 in radiation intensity. To summarize, the dust temperature immediately in front of the shock front $T_{\rm sf}$ does not depend on the temperature in outer environment significantly.

\subsection{Critical Optical Depth \label{sec:critical_optical_depth}}
In order to suppress the isotopic fractionation, it seems that the pre-shock dust temperature, $T_{\rm sf}$ in Eq. (\ref{eq:temperature_shockfront3}), should be lower than $1273 \, {\rm K}$. The requirement gives the upper limit of the optical depth of the pre-shock region above which the isotopic fractionation should occur. Rewriting Eq. (\ref{eq:temperature_shockfront3}), the condition in which the isotopic fractionation is suppressed is given as 
\begin{equation}
\tau_{\rm pre} < \tau_{\rm cr} \equiv \frac{ \strut{\displaystyle{\frac{4 \sigma_{\rm SB} (1273 \, {\rm K})^4}{ \frac{f}{2} \rho_0 v_{\rm s}^3 }}} -2 }{3}.
\end{equation}
We plot the upper limit of the optical depth $\tau_{\rm cr}$ for each shock condition as a function of $f$ in Fig. \ref{fig:tau_cr}. In all shock conditions, large amount of radiation flux returning from the shock front requires the smaller optical depth to suppress the isotopic fractionation. Moreover, the shock condition n11v55 allows the larger optical depth than that of n14v10 because the radiation flux is smaller than that of n14v10. If the optical depth of the pre-shock region is lower than the upper limit in Fig. \ref{fig:tau_cr}, the pre-shock dust temperature does not exceed $1273 \, {\rm K}$ and the isotopic fractionation would be suppressed in the pre-shock region. In this case, in the post-shock region, since the silicate component of dust particles melts quickly due to the gas drag heating, the evaporation of FeS is suppressed and the isotopic fractionation does not take place. To summarize, the upper limit of the pre-shock optical depth $\tau_{\rm pre}$ is about $1-10$ for various shock conditions ($n_0 = 10^{11} - 10^{14} \, {\rm cm^{-3}}$).

\subsection{Different Spatial Scale\label{sec:extended_shock}}
In this study, we assumed $x_{\rm m} = 10^5 \, {\rm km}$, which indicates that the spatial scale of shock-wave heating region is $10^5 \, {\rm km}$. In other words, only dust particles located within $10^5 \, {\rm km}$ from the shock front contribute to the blanket effect, and the thermal radiation of dust particles in outer region runs away effectively. The length of $10^5 \, {\rm km}$ corresponds to about 0.001 AU. On the contrary, if shock waves were generated by the marginally gravitational instability ({\it e.g.,} Boss \& Durisen 2005), the extent of chondrule-forming region would be more larger. However, as we pointed in sec. \ref{sec:radiative_diffusion}, the dust temperature just in front of the shock front is determined by the gas energy flux flowing into the shock front and the pre-shock optical depth $\tau_{\rm pre}$. Since $T_{\rm sf}$ does not explicitly depend on the dimension of the pre-shock region, the spatial scale of the shock waves is not an important factor to consider the rapid heating constraint. In this section, in order to confirm that the optical depth of the pre-shock region $\tau_{\rm pre}$ is a fundamental parameter to determine the pre-shock dust thermal profile, we numerically calculate the case of a different spatial scale ($x_{\rm m} = 10^6 \, {\rm km}$) and the dust-to-gas mass ratio $C_{\rm d} = 0.01$, which results into almost the same $\tau_{\rm pre}$ as the dust-to-gas mass ratio $C_{\rm d} = 0.10$ in the standard spatial scale ($x_{\rm m} = 10^5 \, {\rm km}$). 

Figure \ref{fig:large_scale} shows the dust thermal histories in the pre-shock region for different spatial scale: $x_{\rm m} = 10^6 \, {\rm km}$ and $C_{\rm d} = 0.01$ (top), and $x_{\rm m} = 10^5 \, {\rm km}$ and $C_{\rm d} = 0.10$ (bottom). The shock condition is n14v10 and the dust size distribution is power-law, which result into the pre-shock optical depth of $\tau_{\rm pre} = 10.5$ for top panel and $\tau_{\rm pre} = 14.1$ for bottom panel, respectively. Two dashed lines show the temperature range in which the isotopic fractionation takes plane. It is found that the dust thermal profiles in two cases are very similar in spite that the spatial scales are ten times different. These results indicate that the pre-shock optical depth $\tau_{\rm pre}$ is an fundamental indicator to determine the dust thermal histories in the pre-shock region. 

Strictly speaking, these two dust thermal profiles are not the same exactly because of the slight difference in the pre-shock optical depth $\tau_{\rm pre}$. It is due to the difference of the duration for the dust particles to pass through the pre-shock region; in the top panel, it takes ten times longer than that of the bottom panel. As a result, even larger dust particles can evaporate completely in the top panel. In the top panel, the dust particles whose radii are smaller than $2.51 \, {\rm \mu m}$ evaporate completely in the pre-shock region, on the contrary, in the bottom panel, the dust particles of $0.63 \, {\rm \mu m}$ in radius can survive in this region. As a result, the pre-shock optical depth in the top panel becomes smaller than that of the bottom panel. The difference in the optical properties in the pre-shock region causes the difference in the dust thermal profiles in that region.

\subsection{Cooling Rate Constraint\label{sec:cooling_rate}}
For the chondrule formation, not only the rapid heating constraint, but also the appropriate cooling rate should be satisfied. In our simulations, we can obtain the thermal histories of the dust particles for various radii (peak temperature $T_{\rm peak}$, heating rate $R_{\rm heat}$, and cooling rate $R_{\rm cool}$) as long as the dust particles do not evaporate away in the chondrule-forming region. 

\subsubsection{Appropriate Cooling Rate}
In order to produce chondrules, it is believed that cooling rates after the flash melting event should be appropriate values. However, we discussed the appropriate shock waves only in the view of the rapid heating constraint in the previous section. It is because the cooling rate constraint has not been well determined and could not restrict the flash heating mechanisms so strongly. The data of Jones \& Lofgren (1993), who compared the textures and zoning profiles of olivine grains tens of micrometers across in natural chondrules and experimental analogs, clearly showed that the range of cooling rates experienced by type IIA chondrules was $5 - 100 \, {\rm K/hr}$. For higher cooling rates ($500 - 1000 \, {\rm K/hr}$), olivine textures in these experiments were generally more elongate and skeletal than those grown at the slower cooling rates, and moreover, chemical zoning was rather limited. On the contrary, Wasson \& Rubin (2003) measured overgrowths on low-FeO relict grains contained within almost all of type II chondrules and found that the overgrowths are narrow, in the range of $2 - 12 \, {\rm \mu m}$. Wasson (2004) inferred the cooling rate of chondrules assuming that the overgrowths represent the maximum amount of material that can reach the crystal/melt interface by diffusion in the melt during the cooling interval. Diffusion length can be estimated as $\delta^2 \sim D t_{\rm cool}$, where $\delta$ is the diffusion length, $D$ is the diffusion coefficient, and $t_{\rm cool}$ is the cooling time scale. Thus, the results that the amount of the overgrowths was about 30 times smaller than the grain sizes simulated in order to estimate cooling rates ({\it e.g.,} Jones \& Lofgren 1993) indicate that the cooling rates should be about 900 ($=30^2$) times faster than that suggested by Jones \& Lofgren (1993). Additionally, Yurimoto \& Wasson (2002) modeled the O-isotopic and FeO/(FeO+MgO) gradients in a type II chondrule from the Yamato-81020 CO3.0 chondrite and inferred an extremely fast cooling rate ($10^5 - 10^6 \, {\rm K/hr}$) at high temperatures ($\sim 1900 \, {\rm K}$). To summarize, though the information of the cooling rates is very important for chondrule formation mechanism, the appropriate values have not been well determined yet. We would like to discuss the appropriate shock waves for chondrule formation in the view of the rapid heating and the appropriate cooling rate constraints in the future paper.

\subsubsection{Heating Rate vs. Cooling Rate\label{sec:heating_cooling}}
In order to discuss the chondrule formation, the cooling rate in the phase that the precursor dust particle re-solidifies after the melting phase is an important information. We numerically calculate the thermal histories of the dust particles in the shock-wave heating. We estimate the cooling rate as
\begin{equation}
R_{\rm cool} \equiv \frac{1550 \, {\rm K} - 1400 \, {\rm K}}{\Delta t}, 
\end{equation}
where $\Delta t$ is the duration in which dust temperature decreases from $1550 \, {\rm K}$ to $1400 \, {\rm K}$ in the cooling phase. The results for various shock conditions and the dust models are summarized in Tables \ref{table:summary_a100}$-$\ref{table:summary_a398}. When the dust particle evaporates completely (final dust radius $a_{\rm fin} = 0 \, {\rm \mu m}$), we could not obtain the peak dust temperature $T_{\rm peak}$, the heating rate $R_{\rm heat}$, and the cooling rate $R_{\rm cool}$. We also could not obtain the heating rate $R_{\rm heat}$ if $T_{\rm peak} < 1473 \, {\rm K}$ and the cooling rate $R_{\rm cool}$ if $T_{\rm peak} < 1550 \, {\rm K}$. In those cases, data are not listed in the tables.

Figures \ref{fig:heat_cool_rate_100} - \ref{fig:heat_cool_rate_398} show the plot of the heating rate vs. the cooling rate. Symbols indicate the simulation results and the different symbols represent different gas number density in the pre-shock region; $n_0 = 10^{14} \, {\rm cm^{-3}}$ (circle), $10^{13} \, {\rm cm^{-3}}$ (square), $10^{12} \, {\rm cm^{-3}}$ (triangle), and $10^{11} \, {\rm cm^{-3}}$ (inverse triangle), respectively. The vertical dashed line indicates the lower limit of the heating rate below which the isotopic fractionation would take place. We draw the horizontal lines at $R_{\rm cool} = 5 \, {\rm K \, hr^{-1}}$ and $100 \, {\rm K \, hr^{-1}}$ between which the textures and zoning profiles of olivine grains tens of micrometers across in natural chondrules are well explained (Jones \& Lofgren 1993). We also draw the horizontal line at a higher cooling rate ($R_{\rm cool} \ga 5000 \, {\rm K \, hr^{-1}}$) suggested from the measurement of the overgrowth on low-FeO relict grains contained within almost type II chondrules (Wasson 2004). 

It is found that the numerical results are roughly classified into two categories; the first is the slow heating ($R_{\rm heat} = 10^2 - 10^3 \, {\rm K \, hr^{-1}}$) and slow cooling ($R_{\rm cool} = 5 - 20 \, {\rm K \, hr^{-1}}$), and the second is the rapid heating ($R_{\rm heat} \sim 10^6 \, {\rm K \, hr^{-1}}$) and rapid cooling ($R_{\rm cool} = 10^3 - 10^5 \, {\rm K \, hr^{-1}}$). The former case appears when the gas number density in the pre-shock region is relatively high ($n_0 \ga 10^{13} \, {\rm cm^{-3}}$) and the optical depth is larger than about unity. On the contrary, if the pre-shock gas number density is low ($n_0 \la 10^{12} \, {\rm cm^{-3}}$) or the optical depth is lower than about unity, the thermal histories of the dust particles become the latter case. Namely, the heating rate and the cooling rate become extremely small only when the radiation field is strong enough. The reason can be easily understood as follows. If the mean intensity of the radiation field is strong enough to heat the dust particle up to 1273 K or more in the pre-shock region, the heating rate is determined at the pre-shock region and it becomes small (see sec. \ref{sec:heating_rate}). Moreover, when the radiation field is strong in the pre-shock region, the mean intensity of the radiation field is also strong in the post-shock region (see Fig. \ref{fig:flux}). The strong radiation field keeps the dust particle at high temperature for a long time, so the cooling rate becomes small. 

The rapid heating/cooling case satisfies the rapid heating constraint for suppressing the isotopic fractionation. The cooling rate meets the measurement of the overgrowths on low-FeO relict grains contained within almost all of type II chondrules (Wasson 2004). On the contrary, in the slow heating/cooling case, the isotopic fractionation must take place. However, the cooling rate well matches with the estimation suggested by comparisons of the textures and zoning profiles of olivine grains tens of micrometers across in natural chondrules and experimental analogs (Jones \& Lofgren 1993).

\subsection{Appropriate Shock Mechanism \label{sec:shock_mechanism}}
In order to form chondrules, the shock wave generation mechanism should satisfy at least the rapid heating constraint. We discuss which shock wave generation mechanism is appropriate for chondrule formation in view of the rapid heating constraint. 

\subsubsection{Gravitational Instability}
Recently, Boss \& Durisen (2005) showed that the marginally gravitational instability in the nebula can drive inward spiral shock fronts at asteroidal orbits, sufficient to account for melting of dust particles. Shock waves generated by this mechanism are similar to the shock condition of n14v10. They did not examine if the rapid heating is satisfied or not. If the spatial scale of shock waves is $10^5 \, {\rm km}$, which is the same as the calculation region we assumed in this study, a shock wave with the dust-to-gas mass ratio less than about 0.03 satisfies the rapid heating condition. However, if the spatial scale is $10^6 \, {\rm km}$, which is ten times larger than that of our assumption, it is impossible to suppress the isotopic fractionation even if the dust-to-gas mass ratio is 0.01, which is a standard value of the minimum mass solar nebula model. We clearly showed this conclusion in Fig. \ref{fig:large_scale} for the dust model P01. For a case that all dust particles have the same radius, we can easily derive the same conclusion as follows. The optical depth of the pre-shock region is calculated as $\tau_{\rm pre} \sim (\alpha_{\rm p} + \sigma_{\rm p}) x_{\rm m}$, where $\alpha_{\rm p} + \sigma_{\rm p} = \pi a_{\rm d}^2 (\epsilon_{\rm abs} + \epsilon_{\rm scat}) n_{\rm d}$. Here, we assume that the dust particles have the same radius of $500 \, {\rm \mu m}$. Assuming $\rho_0 = 4 \times 10^{-10} \, {\rm g \, cm^{-3}}$ and the dust-to-gas mass ratio $C_{\rm d} = 0.01$, the number density of dust particles is $n_{\rm d} \sim 2 \times 10^{-9} \, {\rm cm^{-3}}$. For the astronomical silicate with radius of $500 \, {\rm \mu m}$, $\epsilon_{\rm abs} + \epsilon_{\rm scat} \sim 2$. Therefore, substituting $x_{\rm m} = 10^6 \, {\rm km}$, we finally obtain $\tau_{\rm pre} \sim 3$ and it exceeds the upper limit of the optical depth in Fig. \ref{fig:tau_cr}. 

\subsubsection{Planetesimal Bow Shock\label{sec:planetesima_bow_shock}}
Near the nebula midplane, possible energy sources for forming chondrules include bow waves upstream of planetesimals scattered gravitationally into eccentric/inclined orbits by proto-Jupiter (Hood 1998, Weidenschilling {\it et al.} 1998). The gravitational instability shock wave mechanism stated above predicts relatively large-scale sizes ($> 10^4 \, {\rm km}$) for shock-heated regions, while the planetesimal bow shock predicts much smaller scale size ($\sim 10 \, {\rm km}$). Because of the small scale size, the optical depth of the chondrule-forming region will be very small ($\tau_{\rm pre} \sim 10^{-5}$). Since the blanket effect does not work well in such an optically thin environment, the isotopic fractionation would be prevented. 

Even if the optical depth exceeds $\tau_{\rm cr}$, the rapid heating constraint might be satisfied in such small spatial scale. In order to prevent the isotopic fractionation, the precursor dust particles should be heated from below $1273 \, {\rm K}$ to above $1473 \, {\rm K}$ within $\sim (1473 - 1273) / R_{\rm heat} = 72 \, {\rm s}$, where we substitute $R_{\rm heat} = 10^4 \, {\rm K \, hr^{-1}}$. If we take the shock velocity of $v_{\rm s} = 10 \, {\rm km \, s^{-1}}$, the precursor dust particles can travel $720 \, {\rm km}$ within $72 \, {\rm s}$. On the contrary, the spatial scale of the planetesimal bow shock is roughly estimated as $10 \, {\rm km}$, it does not take $72 \, {\rm s}$ to pass through the pre-shock region. Therefore, the isotopic fractionation would not occur significantly. However, since the optically thick environment with such small spatial scale requires a very large dust-to-gas mass ratio ($C_{\rm d} = 100$ or more), this situation might not be present in the protoplanetary disk. 

\subsubsection{Clumpy Cloud Accretion}
Boss \& Graham (1993) suggested that the source of the shocks possibly responsible for chondrule formation was episodic accretion onto the solar nebula of low-mass clumps. The existence of the clumps is suggested by some observational evidences (Bellingham \& Rossano 1980). They considered that the clumps are orbiting above the protoplanetary disk and sometimes obscuring the central star. When the clumps accrete onto the disk surface, the shock waves are generated at the upper region of the disk. Boss \& Graham assumed that the following parameters characterize a ``standard" clump: mass $M_{\rm c} = 10^{22} \, {\rm g}$; density $\rho_{\rm c} = 3\times 10^{-15} \, {\rm g \, cm^{-3}}$; and radius $R_{\rm c} = 9 \times 10^{11} \, {\rm cm}$. Assuming the same parameters as the case of the gravitational instability except for $x_{\rm m} = 10^7 \, {\rm km}$ and $\rho_0 = 4 \times 10^{-13} \, {\rm g \, cm^{-3}}$ (upper region of the protoplanetary disk is low density), we obtained the pre-shock optical depth of $\tau_{\rm pre} \sim 0.3$. The optical depth is low enough to prevent the isotopic fractionation. Therefore, the accretion shock waves seem to match with the rapid heating constraint. However, Tanaka {\it et al.} (1998) carried out the hydrodynamic simulation of the accreting clumps based on the standard clump and they concluded that more than about $10^2$ to $10^3$ times surface densities for the same size of the representative clump are needed to produce chondrules. They also suggested that there are not so many observational studies of the clumpy clouds around the young stellar objects, so more massive clumpy clouds which are sufficient to melt the precursor dust particles might exist. 

\subsubsection{X-ray Flare Induced Shock Wave}
Nakamoto {\it et al.} (2005) proposed a new shock wave generation mechanism: shock waves in the upper region of the solar nebula induced by X-ray flares associated with the young Sun. X-ray flares, common among T Tauri stars (Feigelson {\it et al.} 2002), emit plasma gas, which cools to be a strong neutral gas wind. Hayashi {\it et al.} (1996) suggested that the magnetocentrifugal acceleration (Blandford \& Payne 1982) also plays an important role to produce the wind. Because of the enormous amount of energy released by the X-ray flares and the magnetocentrifugal acceleration, the flares should have some effects on the dynamics and energetics of a protoplanetary disk around the star. Nakamoto {\it et al.} (2005) carried out 2-D MHD numerical simulations of X-ray flares around a central star and expanding magnetic bubles/winds with a disk. They showed that shock waves which are sufficient to melt precursor dust particles can be generated in the upper region of the nebula inside about the asteroid belt. In that case, it is thought that the shock waves are generated in a very wide region at the upper surface of the disk. The neutral gas wind travels in almost parallel with the disk surface much faster than the shock waves propagating vertically in the disk toward the midplane. Therefore, the shock front caused by the neutral gas wind is approximately parallel to the disk surface and the horizontal scale of the shock front would be larger than the scale height of the gas disk. It indicates that the dust particles located at the midplane also can contribute to the blanket effect. As a result, the optical depth perpendicular to the shock front might become much larger than unity. To summarize, there is a possibility that the X-ray flare induced shock waves result in the isotopic fractionation in spite that the chondrule formation occurs in the low gas density region. However, if the shock wave is non-steady, this difficulty may be avoided, because the dust in the midplane can become a large absorber of the radiation flux rather than a reflector and the blanket. This possibility will be discussed in the following subsection.

\subsubsection{Other Possibilities}
We verified four shock generation models as the chondrule formation mechanisms. We found that most of shock waves can be classified into two categories; the rapid heating and the rapid cooling, or the slow heating and the slow cooling (see sec. \ref{sec:heating_cooling}). If supporting an opinion that the slow cooling rate is appropriate for producing chondrules (Jones \& Lofgren 1993), there are few cases in which both of two constraints are satisfied simultaneously (see Fig. \ref{fig:heat_cool_rate_100}$-$\ref{fig:heat_cool_rate_398}). How can we explain the chondrule formation with the rapid heating and the appropriate slow cooling?

One possibility to avoid above difficulties is to consider the non-steady shock wave model. When shock waves are generated in the solar nebula, the shocked gas just behind the shock front emits the line photons and they penetrate into the pre-shock region. Although the radiation heats the dust particles in the pre-shock region, it will take time before the dust temperatures have a steady profile. If the shock waves disappear within a shorter period of time, the dust temperatures would not be as high as the steady profiles. Therefore, the rapid heating constraint might be satisfied even in a relatively optically thick environment. The cooling rate of the precursor dust particles has not been investigated in the framework of the non-steady shock waves, so it is an important issue which should be studied in the future. 

Another possibility is a multiple shock heating scenario. In this study, we assume the porous structures for the precursor dust particles in order to consider the evaporation of sulfur inside the aggregates. In that case, the rapid heating constraint is required in order to prevent the isotopic fractionation. However, if the precursor dust particles have been once melted by some heating process, since the evaporation of sulfur inside the aggregates can hardly occur, the rapid heating might not be required. Based on these discussions, we can consider the multiple shock heating scenario in which the rapid heating constraint and the appropriate slow cooling are both satisfied. In the first heating event, the precursor dust particles should be heated by the shock waves in which the rapid heating constraint is satisfied. At that time, the cooling rate can take any values. After that, in the final heating event, the once melted precursor particles should be heated by the shock waves with the appropriate slow cooling. At that time, the rapid heating is not required because the sulfur cannot evaporate away from inside of the dense particles. In fact, there are some observational evidences of the multiple heating in chondrule formation processes (Jones {\it et al.} 2000), so the above scenario seems to be reasonable. However, it is not clear whether the isotopic fractionation can take place from the once-molten particles or not. The condition of the isotopic fractionation from once-molten particles is also an interesting study.

\subsection{Porosity of Precursor Dust Particles\label{sec:porous}}
For calculating the dust dynamics, we assumed dense structures for the chondrule precursor dust particles (mass density is the same as a value of forsterite). However, it is thought that the precursor dust particles before melting have porous structures, therefore sulfur can evaporate from inside the dust particles and the isotopic fractionation can take place. If the porous structures are considered, a cross section of the dust particle would be larger than the dust model that we adopted in this study. As a result, the optical depth of the pre-shock region $\tau_{\rm pre}$ would be larger, even if the dust model (dust size distribution, dust-to-gas mass ratio) is fixed. In this section, we consider to what extent the porous structures have an effect on our conclusions. 

Figure \ref{fig:porosity} shows schematic pictures of precursor dust particles having various porous structures. The dark gray region indicates a dense region in which the porosity is 0\%. The light gray region indicates a porous region in which the porosity is assumed to be 80\% so that sulfur can evaporate from inside of the porous region. We assume that the porous region is surrounding around the dense core. The volume percent of each region is displayed below each picture. We can obtain the dust radius of each structure from the volume percent of each region and the porosity of the porous region, assuming that the total masses of those particles are the same. We also display the radius and the cross section in units of the values for the case (a), in which all the region of the precursor dust particle is dense. 

If all the region of the precursor dust particles is porous (case (d)), the radius and the cross section are about 1.7 and 2.9 times larger than that of the case (a), respectively. It indicates that the optical depth of the pre-shock region $\tau_{\rm pre}$ becomes about 3 times larger than that obtained in this study. For example, in the case that the shock model is n14v10 and dust model L03 (see Table \ref{table:summary_a251}), we obtained $\tau_{\rm pre} = 0.73$. However, if all dust particles have porous structures like the case (d) in Fig. \ref{fig:porosity}, $\tau_{\rm pre}$ would be about 2.13. The increase of $\tau_{\rm pre}$ would result in that  the dust temperature just in front of the shock front $T_{\rm sf}$ exceeds 1273 K (see Fig. \ref{fig:Tpre2}) and the isotopic fractionation takes place. On the contrary, in a case of the dust model L01, the optical depth of the pre-shock region $\tau_{\rm pre}$ would be about 0.7 if all precursor dust particles are assumed to have porous structures like the case (d) in Fig. \ref{fig:porosity}. This expected value of $\tau_{\rm pre}$ seems not to be so large that the isotopic fractionation would not occur. To summarize, we think that the discussion about the upper limit of $\tau_{\rm pre}$ for preventing the isotopic fractionation in secs. \ref{sec:radiative_diffusion} and \ref{sec:critical_optical_depth} is not affected significantly by considering the porous structures because the dust temperature just in front of the shock front $T_{\rm sf}$ mainly depends only on $\tau_{\rm pre}$ and the net flux of the radiation ${\cal F}$ (see Section \ref{sec:radiative_diffusion}). However, the corresponding dust model (dust size distribution, dust-to-gas mass ratio) for a certain value of $\tau_{\rm pre}$ would be affected. 

Finally, we notice that the structures of chondrule precursor dust particles have not been well investigated. Although we assumed the porosity of the porous region as 80\% for sulfur to evaporate from inside, a value of the porosity is unknown. Moreover, it is not certain whether all region of the precursor dust particles should be porous or not. We can imagine the mixed structure like the case (b) in Fig. \ref{fig:porosity}, which has the dense core surrounded by the porous mantle. Of course, since sulfur would be able to evaporate from inside of the porous mantle, it must be heated rapidly enough to prevent the isotopic fractionation. In the case (b), the cross section is about 1.5 times larger than that of the case (a). Therefore, although the internal structure of precursor dust particle would affect the optical properties of the pre-shock region, it is a very complex problem. If we take into account the internal structure of the dust particle into our model, it will require a large amount of parameter sets which characterize the dust structure. It is beyond the scope of this paper to investigate for all of the parameter sets.

\subsection{Plane-parallel Geometry\label{sec:validity_plane-parallel}}
We assumed the one-dimensional, plane-parallel, and steady shock structures in this study. In these situations, the radiation flux ${\cal F}$ should be constant in the pre-shock region because there are not any radiative sources or sinks (see Appendix \ref{appen:constancy_radiation_flux}). It is clearly shown in our numerical results (see Fig. \ref{fig:flux}). It indicates that the radiative heating takes place everywhere in the pre-shock region, even at the upstream boundary of the computational domain ($x = - x_{\rm m}$). The dust temperature at the boundary can be estimated from Eq. (\ref{eq:temperature_shockfront3}) by substituting $\tau_{\rm pre} = 0$. We obtain $T_{\rm d} |_{x=-x_{\rm m}} = 969 \, {\rm K}$ for $f=0.5$, $\rho_0 = 10^{-10} \, {\rm g \, cm^{-3}}$, and $v_{\rm s} = 10 \, {\rm km \, s^{-1}}$. On the contrary, we assume that the ambient temperature for the gas and the dust particles is $300 \, {\rm K}$. How can we recognize this temperature jump?

The answer is the three-dimensional effect of the chondrule-forming region. It is considered that the one-dimensional plane-parallel approximation would be adoptable near the shock front. In this region, the radiation would travel perpendicularly to the shock front and the radiative flux ${\cal F}$ should be constant. However, far from the shock front, which is roughly estimated by the spatial extent of the shock front, the plane-parallel assumption breaks down. The radiation can diffuse away not only perpendicularly to the shock front but also to the other directions, therefore, the effect of the radiative heating becomes weak rapidly as it keeps away from the shock front. Finally, the dust temperature becomes as the same as the ambient temperature. 

If the thermal histories of the dust particles below about $1000 \, {\rm K}$ are in interest, it would be needed to develop the two- or three-dimensional shock-wave heating model. In the case of the planetesimal bow shock, some studies about the multi-dimensional shock-wave heating have been carried out (Hood 1998, Weidenschilling {\it et al.} 1998), in which the planetesimals are relatively small ($\sim 10 - 100 \, {\rm km}$) and the multi-dimensional effect should be considered. In our study, we consider relatively large scale shock waves ($\sim 10^5 \, {\rm km}$ or more). Moreover, the thermal histories of the dust particles below $1000 \, {\rm K}$ do not play an important role for our purpose. Therefore, we do not treat the thermal evolutions of the dust particles before they enter the one-dimensional plane-parallel region.

\subsection{How Strong is Heating Constraint?\label{sec:validity_rapid_heating}}
In this study, we discussed the appropriate shock conditions and the dust models for chondrule formation with the rapid heating constraint, derived from the measurements of the isotopic composition of sulfur (Tachibana \& Huss 2005). However, as commented in their abstract, this constraint is based on the assumption that the troilite grains within 15 ferromagnesian chondrules they measured are primary. They considered that some troilite grains have survived chondrule-forming high-temperature events because they are located inside metal spherules within chondrules. According to them, such an occurrence is unlikely to be formed by secondary sulfidization processes in the solar nebula or on parent bodies. 

If those troilite grains are not primary, the rapid heating constraint would not be required necessarily. In our study, we found that the rapid heating ($R_{\rm heat} \sim 10^6 \, {\rm K \, hr^{-1}}$) does not seem to be compatible with the slow cooling ($R_{\rm cool} \sim 5 - 20 \, {\rm K \, hr^{-1}}$), which is considered by many researchers to be appropriate for chondrule formation (e.g., Jones \& Lofgren 1993). However, we do not mean that we should adopt the rapid heating constraint rather than the slow cooling rate. We just investigated the thermal evolutions of the dust particles in the shock-wave heating for various situations by using a more realistic physical model. The rapid heating constraint is not needed to be assumed in order to derive our numerical results. We think that more experiments and observations are required for elucidating the riddles of chondrule formation. We believe that our simulations give a great insight of the shock-wave heating model for chondrule formation to readers.

\section{CONCLUSIONS}
We numerically simulated the shock-wave heating for chondrule formation taking into account the radiation transfer of the line emission of gas molecules and the dust thermal continuum emission. Regarding the line cooling due to the gas molecules, we estimated the cooling rate using the photon escape probability method in order to reflect the column density of gas molecules. We focused the dust thermal histories in the pre-shock region, especially the heating rate of chondrules $R_{\rm heat}$, which has to be large enough to suppress the isotopic fractionation. In order to clear the condition in which the rapid heating constraint is satisfied, we performed simulations for various shock conditions, the pre-shock gas number density ($n_0 = 10^{11} - 10^{14} \, {\rm cm^{-3}}$) and the shock velocity ($v_{\rm s} = 10 - 55 \, {\rm km \, s^{-1}}$), and the dust models, the initial dust size distribution (the power-law size distribution and the lognormal one) and the dust-to-gas mass ratio ($C_{\rm d} = 0.01$, $0.03$, and $0.10$). We found the following results: 
\begin{enumerate}
\item The line emission can remove the post-shock gas thermal energy away even if the pre-shock gas number density is relatively high ($n_0 = 10^{14} \, {\rm cm^{-3}}$). Therefore, the line emission plays an important role as the gas cooling mechanism. Moreover, it indicates that it is also an important radiation source term. 
\item When the optical depth of the pre-shock region increases, it results into the higher dust temperatures in the pre-shock region. If the dust temperatures just in front of the shock front exceed $1273 \, {\rm K}$, the heating rate of the dust particles in a temperature range of $1273 - 1473 \, {\rm K}$ is too slow to prevent the isotopic fractionation. On the contrary, the dust temperatures just in front of the shock front are lower than $1273 \, {\rm K}$, the dust particles are heated by the gas frictional heating in the post-shock region so rapidly that the isotopic fractionation is prevented. Therefore, the condition to prevent the isotopic fractionation is that the dust temperatures just in front of the shock front do not exceed $1273 \, {\rm K}$. 
\item We analytically derived the dust temperature just in front of the shock front using the theory of the radiative diffusion. The analytic solution well explains the results of our numerical simulations for optically thick case (optical depth of the pre-shock region $\tau_{\rm pre} \ga 1$). 
\item There is the upper limit of the optical depth of the pre-shock region above which the isotopic fractionation will occur. The value of the upper limit depends on the shock condition. For the low-velocity and high-density shock waves, the upper limit is about unity. For the high-velocity and low-density shock waves, it is about 10. 
\item The fundamental factors to determine the pre-shock dust thermal histories are the gas energy flux flowing into the shock front and the optical depth of the pre-shock region. Even if the spacial dimension of the shock-wave heating region is different, the dust temperatures just in front of the shock front become almost the same as long as the pre-shock regions take the similar values of the optical depths. 
\item We also evaluated the cooling rate of the precursor dust particles $R_{\rm cool}$ at a crystallization temperature. It was found that the dust thermal histories obtained by our simulations are roughly classified into two categories: the first is the slow heating ($R_{\rm heat} = 10^2 - 10^3 \, {\rm K \, hr^{-1}}$) and slow cooling ($R_{\rm cool} = 5 - 20 \, {\rm K \, hr^{-1}}$), and the second is the rapid heating ($R_{\rm heat} \sim 10^6 \, {\rm K \, hr^{-1}}$) and the rapid cooling ($R_{\rm cool} = 10^3 - 10^5 \, {\rm K \, hr^{-1}}$). If supporting an opinion that the slow cooling rate is appropriate for producing chondrules, it seems difficult to meet two conditions simultaneously in the framework of a one-dimensional, plane-parallel, and steady shock wave model. In order to satisfy the rapid heating constraint and the appropriate slow cooling constraint simultaneously, a non-steady shock wave model or a multiple heating scenario might be needed. 
\end{enumerate}




\acknowledgments

We are grateful to Dr. S. Tachibana for useful discussion on observations of  isotopic fractionation and pre-shock heating rates. Helpful discussions with Prof. Umemura are gratefully acknowledged. Thanks are due to our colleagues with whom we have discussed this problem. HM was supported by the Research Fellowship of Japan Society for the Promotion of Science for Young Scientists. TN was supported by Grant-in-Aid for Young Scientists (B) (14740284) of Japan Society for the Promotion of Science. 





\appendix

\section{Radiative Transfer Equation}
One of the purposes of this study is to develop the numerical code of shock-wave heating model taking into account the dust thermal continuum emission and the line emission of gas molecules. However, to solve the transfer of the line emission with no approximation needs a huge amount of computational time. So, we approximate the line transfer problem as follows. 

Assuming the isotropic scattering, the frequency-dependent radiation transfer equation taking into account absorption and scattering by the gas molecules and dust particles is given as (e.g., Rybicki \& Lightman 1979)
\begin{equation}
\frac{d{\cal I}_\nu}{dx} = - (\alpha_{\nu, {\rm g}} + \alpha_{\nu, {\rm d}} + \sigma_{\nu, {\rm g}} + \sigma_{\nu, {\rm d}} ) {\cal I}_\nu + j_{\nu, {\rm d}} + j_{\nu, {\rm g}} + ( \sigma_{\nu, {\rm g}} + \sigma_{\nu, {\rm d}} ) {\cal J}_\nu, 
\end{equation}
where $\alpha_\nu$, $\sigma_\nu$, and $j_\nu$ are the absorption coefficient, the scattering coefficient, and the emission coefficient, respectively. The subscripts ``g" and ``d" mean the gas and dust particles, respectively. The specific intensity is ${\cal I}_\nu$ and the mean intensity is ${\cal J}_\nu$ ($\equiv (4 \pi)^{-1} \int {\cal I}_\nu d\Omega$). The first term of the right hand side means the extinction by absorption and scattering, the second and third terms are the source term due to the dust thermal continuum emission and the line emission, and fourth term is the source term due to the scattering. We assume that the effects of the absorption and the scattering of line emissions by the gas molecules can be included into the emission coefficient $j_{\nu, {\rm g}}$ and we denote the {\it net} emission coefficient  as $j_{\nu, {\rm g}}'$. Using the notation, the radiation transfer equation can be rewritten as 
\begin{equation}
\frac{d{\cal I}_\nu}{dx} = - ( \alpha_{\nu, {\rm d}} + \sigma_{\nu, {\rm d}} ) {\cal I}_\nu + j_{\nu, {\rm d}} + j_{\nu, {\rm g}}' + \sigma_{\nu, {\rm d}} {\cal J}_\nu . 
\label{appendix:radiation_transfer}
\end{equation}

Integrating Eq. (\ref{appendix:radiation_transfer}) over $\nu$, we obtain
\begin{equation}
\frac{d{\cal I}}{dx} = - ( \tilde{\alpha} + \tilde{\sigma} ) {\cal I} + j_{\rm d} + j_{\rm g}' + \bar{\sigma} {\cal J}, 
\label{eq:radtr03}
\end{equation}
where $\tilde{\alpha} \equiv \int \alpha_{\nu, {\rm d}} {\cal I}_\nu d\nu / \int {\cal I}_\nu d\nu$, $\tilde{\sigma} \equiv \int \sigma_{\nu, {\rm d}} {\cal I}_\nu d\nu / \int {\cal I}_\nu d\nu$, and $\bar{\sigma} \equiv \int \sigma_{\nu, {\rm d}} {\cal J}_\nu d\nu / \int {\cal J}_\nu d\nu$ are the frequency-averaged absorption/scattering coefficients. Here, we replace these frequency-avaraged absorption/scattering coefficients to the Planck mean values. We think that it is an appropriate approximation because the line center frequency is similar to the peak frequency of the dust thermal continuum emission. Finally, we can rewrite Eq.(\ref{eq:radtr03}) into a simpler form 
\begin{equation}
\frac{d{\cal I}}{dx} = -(\alpha_{\rm p} + \sigma_{\rm p}) ( {\cal I} + {\cal S} ), 
\label{eq:radtr04}
\end{equation}
where
\begin{equation}
{\cal S} \equiv \frac{ j_{\rm d} + j_{\rm g}' + \sigma_{\rm p} {\cal J} }{ \alpha_{\rm p} + \sigma_{\rm p} }. 
\label{eq:source_function}
\end{equation}

\section{Escape Probability Method\label{appen:escape_probability_molecule}}

In the shock-wave heating model, the optical depth effects are important in determining the net cooling rate by line emissions of gas molecules. In order to evaluate the net cooling rate, we adopt the fitting formula derived by Neufeld \& Kaufman (1993), in which they numerically solved the equations of statistical equilibrium for the rotational/vibrational level populations using an escape probability method. 

The equations of statistical equilibrium is given as
\begin{equation}
\sum_j f_j (A_{ji} \beta_{ji} + C_{ji} ) - f_i \sum_j (A_{ij} \beta_{ij} + C_{ij}), 
\label{eq:statistic_equilibrium}
\end{equation}
where $f_i$ is the fractional population in state $i$, $A_{ij}$ and $C_{ij}$ are the spontaneous radiative rate and collisional rate for transitions from state $i$ to state $j$, and $\beta_{ij}$ is an angle-averaged escape probability and given by
\begin{equation}
\beta_{ij} \simeq \frac{ 1 }{ 1 + 3 \tau_{\rm S} }, 
\label{eq:escape_probability}
\end{equation}
where the Sobolev optical depth $\tau_{\rm S}$ is given by
\begin{equation}
\tau_{\rm S} = \frac{ hcn({\rm M}) }{ 4 \pi | dv_z / dz | } (f_j B_{ji} - f_i B_{ij} ), 
\label{eq:sobolev_optical_depth}
\end{equation}
where $n({\rm M})$ is the volume density of coolant molecules, $B_{ji}$ and $B_{ij}$ are the absorption and stimulated emission coefficients, and $dv_z / dz$ is the local velocity gradient. Eq. (\ref{eq:sobolev_optical_depth}) is exact in the limits of small and large $\tau_{\rm S}$, and deviates from the exact expression of Hummer \& Rybicki (1982) by at most 15\%. Solving the equations of statistical equilibrium is made more difficult by the fact that the escape probabilities for the emitted line photons are themselves a function of the level populations. Neufeld \& Kaufman numerically solved those equations including a large number of rotational states and obtained the net cooling rates over a wide range of physical conditions. 

Neufeld \& Kaufman (1993) found a convenient analytic fit to express their numerical solutions. The net cooling rate $\Lambda$ may be conveniently expressed as a rate coefficient, $L$, defined such that the cooling power per unit volume due to collisions of H$_2$ and species M is given by $\Lambda = L n({\rm H_2}) n ({\rm M})$, where $n({\rm H_2})$ and $n({\rm M})$ are the H$_2$ and coolant particle densities. The cooling rate coefficient $L$ has units of erg cm$^3$ s$^{-1}$. The analytic fit for $L$ is written as
\begin{equation}
\frac{1}{L} = \frac{1}{L_0} + \frac{ n({\rm H_2}) }{ L_{\rm LTE} } + \frac{ 1 }{ L_0 } \bigg[ \frac{ n({\rm H_2}) }{ n_{1/2} } \bigg]^\alpha \bigg( 1 - \frac{ n_{1/2} L_0 }{ L_{\rm LTE} } \bigg),
\label{eq:analytic_fit_escape_probability}
\end{equation}
where $L_0$ is a function of temperature alone and $L_{\rm LTE}$, $n_{1/2}$, $\alpha$ are functions of both temperature and $\tilde{N}({\rm M})$, and $\tilde{N}({\rm M})$ is the optical depth parameter. In a case that the emitting point is located at the center of a static plane-parallel slab of thickness $d$, $\tilde{N}({\rm M})$ is expressed as
\begin{equation}
\tilde{N}({\rm M}) = \frac{ n({\rm M}) d }{ \Delta v }, 
\label{eq:optical_depth_parameter}
\end{equation}
where $\Delta v$ is the velocity dispersion of the coolant molecules. For vibrational cooling, Neufeld \& Kaufman recommended a two-parameter fit to the net cooling rate which includes only the first two terms in Eq. (\ref{eq:analytic_fit_escape_probability}). The values of $L_0$, $L_{\rm LTE}$, $n_{1/2}$, and $\alpha$ are listed in Tables 2-5 of Neufeld \& Kaufman (1993).

\subsection{Net Cooling Rate\label{sec:net_cooling_rate}}
We consider line photons emitted at $x=x_A$ in the post-shock region (see Fig. \ref{fig:escape_prob}). We assume that the photons which achieve at $x = x_{\rm m}$ without being absorbed can remove the gas thermal energy to outside. Similarly, it is assumed that the gas molecules in the pre-shock region do not absorb the line photons emitted in the post-shock region because of the doppler shift; the relative velocity between the pre- and post-shock gases is larger than the velocity dispersion of gas molecules. We define the optical depth parameter of the upstream (downstream) side of the emitting point, $\tilde{N}^{up}$ ($\tilde{N}^{down}$), as
\begin{equation}
\tilde{N}^{up}({\rm M}) = 2 \int_0^{x_{\rm emit}} \frac{ n({\rm M}) }{ \Delta v } dx, ~~~
\tilde{N}^{down}({\rm M}) = 2 \int_{x_{\rm emit}}^{x_{\rm m}} \frac{ n({\rm M}) }{ \Delta v } dx. 
\label{eq:optical_depth_parameter_cooling}
\end{equation}
We obtain the rate coefficient for each direction, $L^{up}$ and $L^{down}$, by substituting $\tilde{N}^{up}$ and $\tilde{N}^{down}$ into Eq. (\ref{eq:analytic_fit_escape_probability}). Finally, we obtain the cooling rate for each direction as
\begin{equation}
\Lambda^{up} = L^{up} n({\rm H_2}) n({\rm M}), ~~~
\Lambda^{down} = L^{down} n({\rm H_2}) n({\rm M}). 
\label{eq:cooling_rate_each_direction}
\end{equation}
The net cooling rate is given as 
\begin{equation}
\Lambda' = \frac{\Lambda^{up} + \Lambda^{down}}{2}. 
\label{eq:net_cooling_rate}
\end{equation}

\subsection{Emission Coefficient\label{sec:emission_coefficient}}
In general, assuming an isotropic emitter, we can write the emission coefficient for the radiation transfer equation (Eq. \ref{eq:source_function}) as $j_{\rm g} = \Lambda/(4 \pi)$. So, in this study, we obtain the emission coefficient as
\begin{equation}
j_{\rm g}' = {
\Lambda^{up} / (4 \pi), ~~~~~~ {\rm for} ~ x_A > x_B, 
\atopwithdelims\{. 
\Lambda^{down} / (4 \pi), ~~~ {\rm for} ~ x_A < x_B, 
}
\label{eq:emission_coefficient_line}
\end{equation}
where $\Lambda^{up}$ and $\Lambda^{down}$ are the net cooling rates for the direction of upstream and downstream, respectively (see Eq. \ref{eq:cooling_rate_each_direction}). 

However, it would be an underestimation if we give the emission coefficient as $\Lambda' / (4 \pi)$ because we assume that only the photons which can escape from the post-shock region contribute to the net cooling rate for the gas (see Eq. \ref{eq:net_cooling_rate}). Let's obtain the mean intensity ${\cal J}$ and net flux ${\cal F}$ at $x=x_B$ in Fig. \ref{fig:escape_prob}. The net cooling rate $\Lambda'$ at $x = x_A$ does not include the line emissions which are absorbed by gas molecules located from $x = 0$ to $x = x_A$. On the contrary, when we calculate the radiation field at $x = x_B$, we should consider the absorption between $x = x_B$ and $x = x_A$. Therefore, the mean intensity obtained by our model might be weaker than that expected in a real situation. 

How much different is the mean intensity ${\cal J}$ between our model and the real situation? We think that the difference is not significant in the dust temperatures just in front of the shock front. Assuming the isotropic radiation field in the pre-shock region and the optically thick environment, the dust temperature just in front of the shock front $T_{\rm sf}$ is given by Eq. (\ref{eq:temperature_shockfront3}), where $f$ indicates the net flux of the radiation field returning from the post- to pre-shock region, in the unit of the gas energy flux flowing into the shock front. For example, when all the gas energy flux is converted into the radiation energy at the post-shock region and it returns toward the pre-shock region, we obtain $f = 1$. Therefore, in Fig. \ref{fig:Tpre2}, a curve labeled by ``$f=1$" gives the upper limit of $T_{\rm sf}$ under the assumption of the isotropic radiation field. Although the dust temperature in a case of the dust model L01 exceeds the upper limit, it would be due to an anisotropic radiation field resulting from the optically thin environment. In an optically thick environment ($\tau \ga 1$), since the radiation field tends to be isotropic, the dust temperature just in front of the shock front would be equal or less than the upper limit. The upper limit indicates that when the pre-shock optical depth $\tau_{\rm pre}$ is less than about unity, $T_{\rm sf}$ becomes less than $1273 \, {\rm K}$ and therefore the isotopic fractionation would not take place. To summarize, we might underestimate the mean intensity of the radiation field in our model, however, our conclusion that the critical optical depth for the isotopic fractionation is about $1-2$ does not change significantly. 

However, there is a possibility that the anisotropy of the pre-shock radiation field affects the dust thermal histories in the pre-shock region slightly. Therefore, in order to discuss the dust thermal histories with higher accuracy, we need to develop a new radiation transfer code for the line emissions.  We would like to solve the problem in the future.

\section{Lyman $\alpha$ Cooling}
Regarding the Ly$\alpha$ cooling, we neglect the absorption by other hydrogen atoms because the optical depth for the Ly$\alpha$ photon is small. The absorption coefficient for the Ly$\alpha$ emission is given as $\alpha_\nu = \frac{h\nu}{4\pi} n_1 B_{12} \phi(\nu)$, where $h$ is the Planck constant, $n_1$ is the number density of hydrogen atom in ground state ($n=1$), $B_{12}$ is the Einstein B-constant, and $\phi (\nu)$ is the line profile function (Rybicki \& Lightman 1979). Einstein B-constant is given from the Einstein A-constant by Einstein relation. The line profile function has the maximum value of $\sim 1/\Delta \nu$ at the line center, where $\Delta \nu$ is the Doppler broadening in frequency. In the shock condition n11v55, $n_1 \sim 10^9 \, {\rm cm^{-3}}$ and $\Delta \nu \sim 10^{-4} \nu$ behind the shock front, so we obtain $\alpha_\nu \sim 10^{-5} \, {\rm cm^{-1}}$. It indicates that Ly$\alpha$ photons are not absorbed until they achieve at about $10^5 \, {\rm cm}$ behind the shock front. However, our calculation results show that the gas cools until it reaches that point for the shock condition n11v55. Moreover, the gas decelerated as it cools in the post-shock region. The relative velocity between the shocked gas immediately behind the shock front and the cooled gas far from the shock front is about $10^6 \, {\rm cm \, s^{-1}}$ and it is comparable to the velocity dispersion of the gas molecules. Therefore, the large relative velocity does not allow the cooled gas to absorb Ly$\alpha$ photons emitted immediately behind the shock front. 

The cooling rate due to Ly$\alpha$ emission is given by Spitzer (1978) as
\begin{equation}
\Lambda_{\rm Ly\alpha} = 7.3\times10^{-19} n_e n_{\rm H} \exp (-118,400/T) \, {\rm erg \, cm^{-3} \, s^{-1}}, 
\end{equation}
where $n_e$ and $n_{\rm H}$ are number densities of electron and hydrogen atom in unit ${\rm cm^{-3}}$, respectively, and $T$ is the gas temperature in unit K.

\section{Constancy of Radiation Flux in Pre-shock Region\label{appen:constancy_radiation_flux}}

\subsection{Reason for Constancy}
The radiation flux ${\cal F}$ changing with the distance from the shock front indicates that there are some energy sources (for $d {\cal F} / dx > 0$) or sinks (for $d{\cal F}/dx < 0$) of the radiation. We can consider following two possibilities as the radiation sources/sinks; (1) the medium (gas and/or dust particles) continues to emit or absorb the radiation energy and (2) the medium absorbs the radiation energy and then emits it at other places. 

Let us consider the first possibility. In the post-shock region, it can occur. Behind the shock front, there are the hot gas and dust particles which are heated by a shock wave. They emit the radiation, then cool and flow downstream, however, the new medium which is heated at the shock front continues to be supplied. Therefore, there are eternal radiation sources behind the shock front (for the steady state). On the contrary, in the pre-shock region, the gas and the dust particles cannot be the radiation energy sources. In this region, the main energy source is the radiation coming from the post-shock region. The gas and the dust particles in the pre-shock region just emit the radiation energy which is equal to the radiation energy that they absorbed. It indicates that they can neither produce nor consume the radiation energy inside themselves. Therefore, in the pre-shock region, the first possibility will vanish. 

The second possibility will also vanish if we consider the non-relativistic shock velocity. The radiation energy coming from the post-shock region diffuses into the pre-shock region at the effective speed of $\sim c / 3\tau_{\rm pre}$, where $c$ is the light velocity in the vacuum and $\tau_{\rm pre}$ is the optical depth of the pre-shock region (e.g., Mihalas \& Mihalas 1999, p.353). Considering $\tau_{\rm pre} \sim 100$, the light diffusion velocity is estimated as $c / 3\tau \sim 10^3 \, {\rm km \, s^{-1}}$. Since this is much larger than the shock velocity of about $7 \, {\rm km \, s^{-1}}$ adopted by Desch \& Connolly (2002) as the canonical case, the radiation energy can escape from the pre-shock region so fast that the gas and the dust particles cannot transport the radiation energy. Therefore, the second possibility will also vanish. 

To summarize, the radiation flux in the pre-shock region should be constant. It indicates that the radiative heating takes place even far from the shock front.

\subsection{Reason for Non-Constancy}
In Fig. \ref{fig:flux}, the radiative flux ${\cal F}$ of model L10 does not seem to be constant in the pre-shock region. We thought that the reason is the shortage of the iteration to converge the mean intensity in the computational domain. In our model, we solve the gas/dust dynamics and the radiation transfer by turns until the mean intensity does not change more than 1\% (see sec. \ref{sec:overview_model}). However, this condition does not guarantee the complete convergence. 

Consider that the number of iteration which is required to satisfy above converge condition, $N_{\rm ite}$, exceeds 100. It indicates that the convergence of the mean intensity is very slow. If the iteration process has been continued more and more, the mean intensity would change near 1\% per iteration for a while. Repeating the iteration further 100 times, the mean intensity might change $\sim 1\% \times 100 \sim 100\%$! Therefore, roughly speaking, the above converge condition is not enough for the complete convergence when $N_{\rm ite} \ga 100$. The numerical results for models of L03 and L01 would have converged enough because $N_{\rm ite} < 100$. However, for model of L10, the numerical result would not have converged completely because $N_{\rm ite} > 100$. 

However, we do not think that we need more iteration in order to converge completely. In model of L10, the isotopic fractionation takes place because the dust temperature just in front of the shock front $T_{\rm sf}$ exceeds $1273 \, {\rm K}$. If we have carried out the more iteration, $T_{\rm sf}$ would increase more and more. This does not change the result that the isotopic fractionation occurs. Therefore, we think that our converge condition is meaningful to obtain our conclusions. 

\begin{figure}
\epsscale{.70}
\plotone{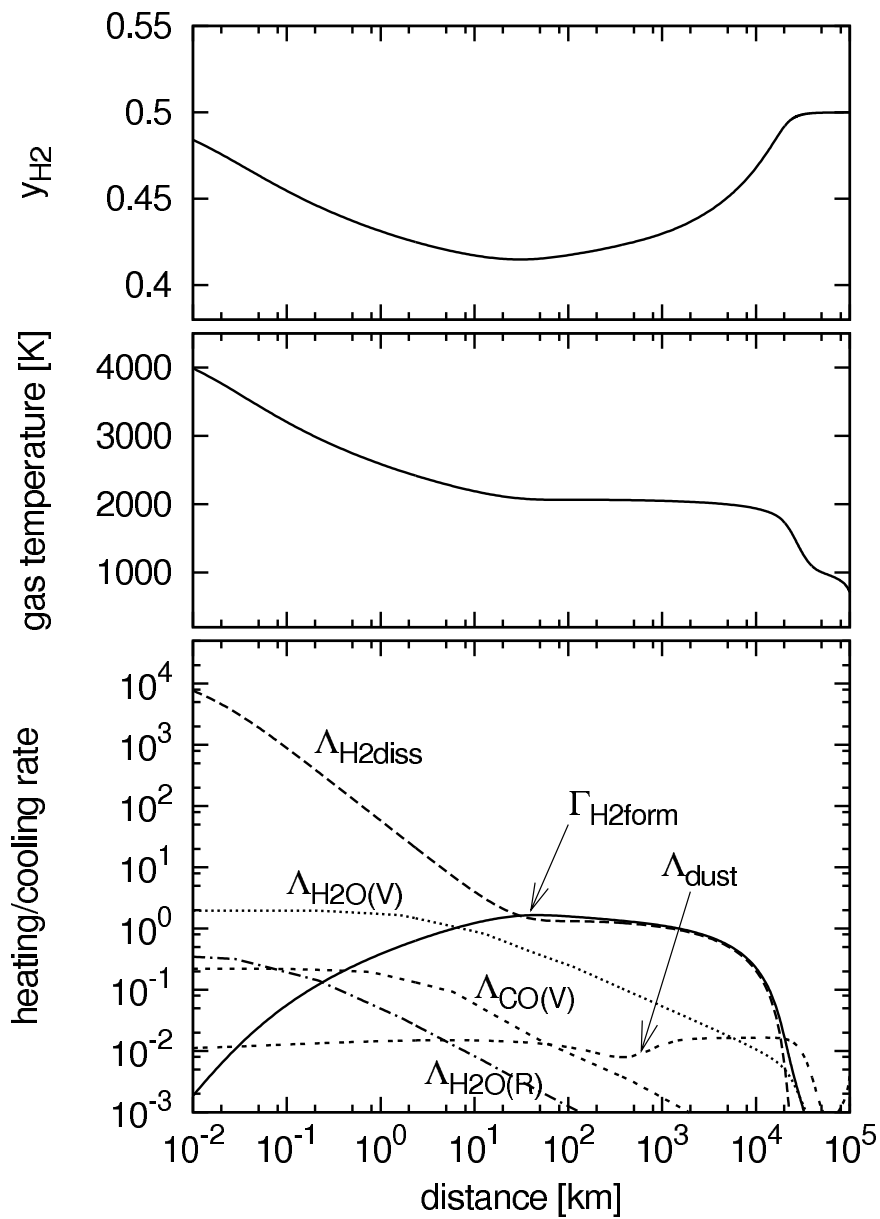}
\caption{Abundance of hydrogen molecule ($y_{\rm H_2}$), gas temperature ($T_{\rm g}$), and heating/cooling rates for gas in the post-shock region are plotted as a function of distance from the shock front. The shock condition is n14v10 and the dust model is L01. The unit of the heating/cooling rates is ${\rm erg \, cm^{-3} \, s^{-1}}$. \label{fig:heatcoolrate_gas}}
\end{figure}

\begin{figure}
\epsscale{.70}
\plotone{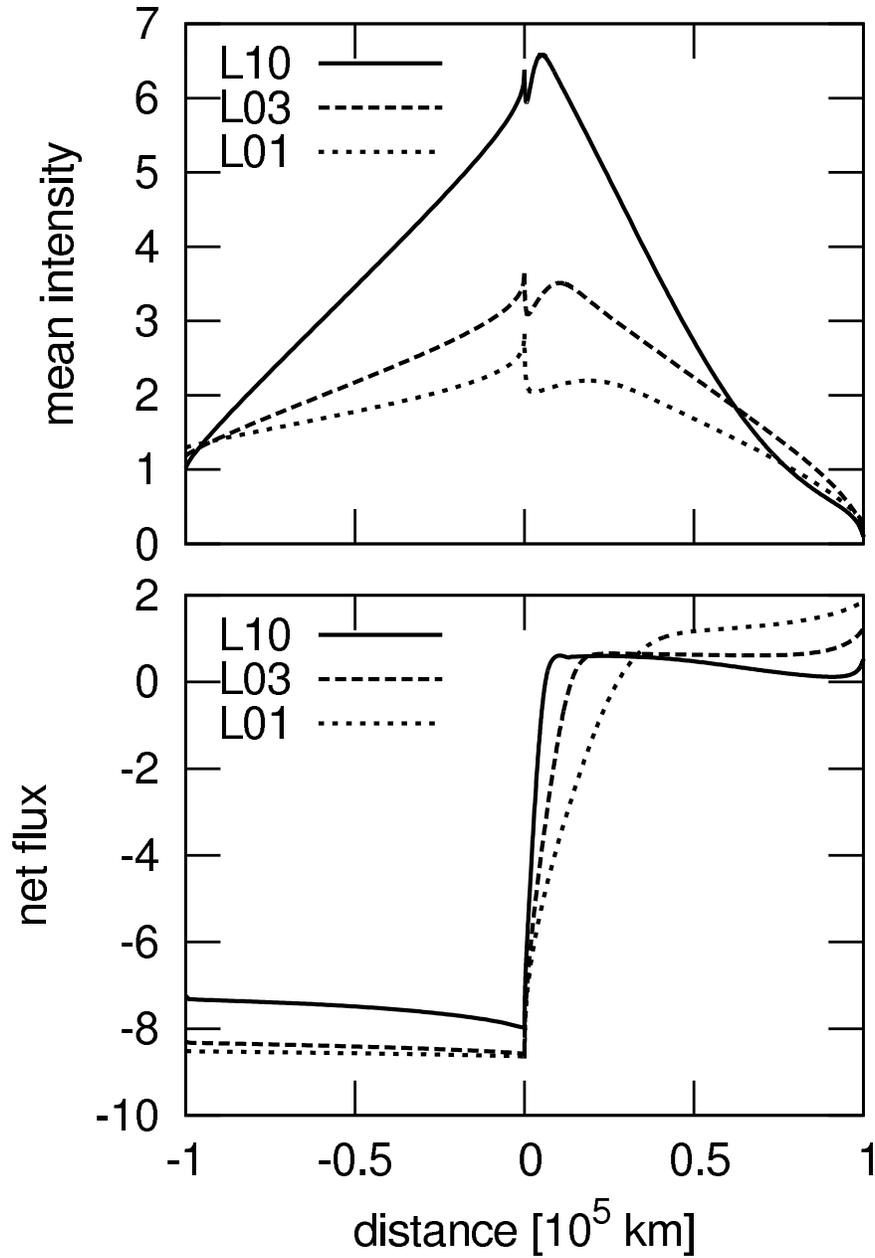}
\caption{Radiation fields in the chondrule forming-region for various dust models are plotted as a function of distance from the shock front. Shock condition is n14v10. Top panel is the mean intensity in unit of $10^7 \, {\rm erg \, cm^{-2} \, ster^{-1} \, s^{-1}}$ and bottom panel is the radiation flux in unit of $10^7 \, {\rm erg \, cm^{-2} \, s^{-1}}$. The negative flux means the net radiation flux flows toward upstream. \label{fig:flux}}
\end{figure}

\begin{figure}
\epsscale{0.9}
\plotone{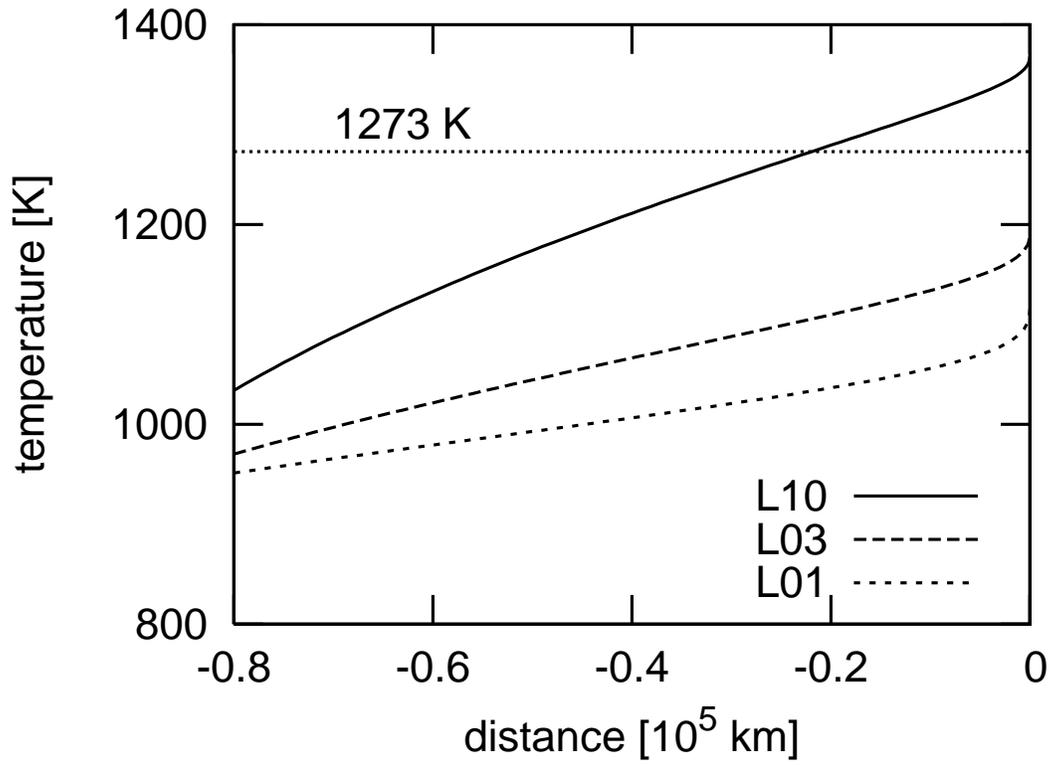}
\caption{Dust thermal histories in the pre-shock region are plotted as a function of distance from shock front. Shock condition is n14v10, and dust models are L10, L03, and L01. A horizontal dotted line indicates $1273 \, {\rm K}$, above which the isotopic fractionation takes place until dust temperature exceeds $1473 \, {\rm K}$. \label{fig:dust_temp}}
\end{figure}

\begin{figure}
\epsscale{.70}
\plotone{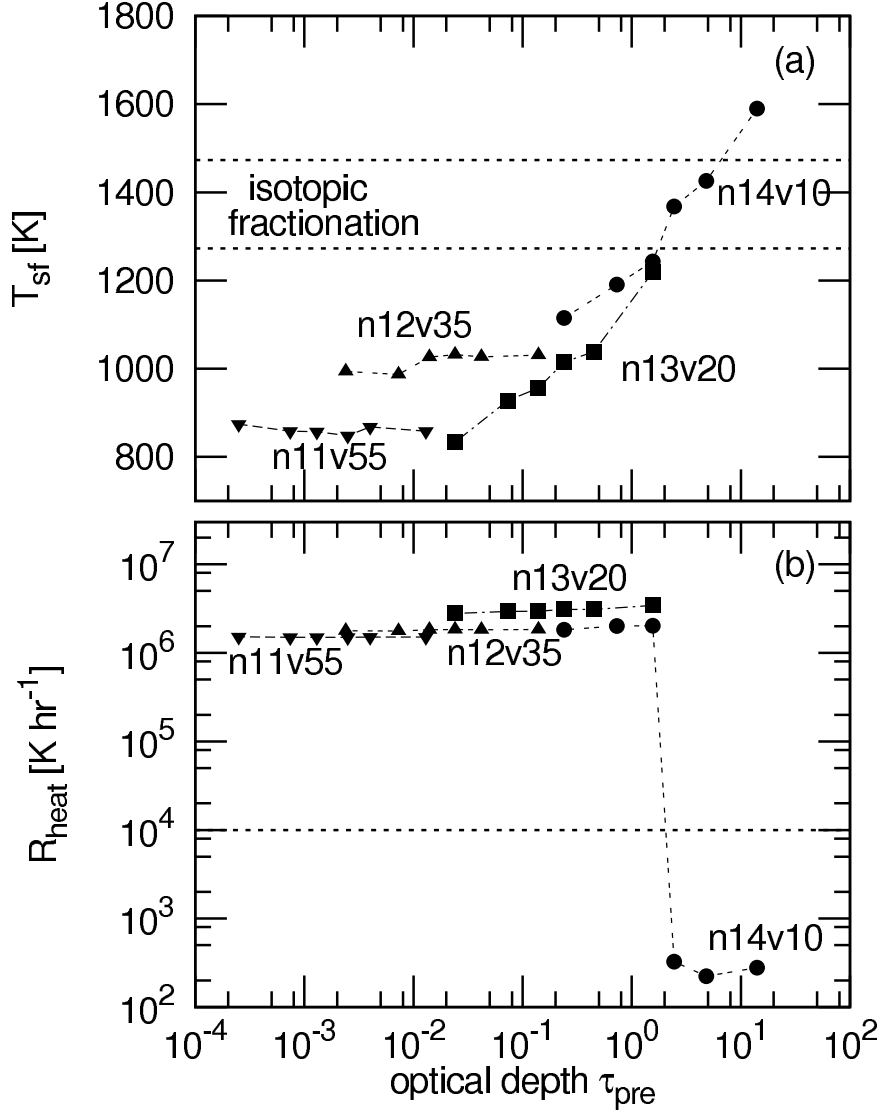}
\caption{Results of dust thermal histories in the pre-shock region are displayed as a function of the optical depth of the pre-shock region: (a) dust temperatures immediately in front of the shock front, which are elevated by heating due to the ambient radiation field, and (b) the heating rate of chondrules in a temperature range of $1273 - 1473 \, {\rm K}$, which is the region surrounded by two dashed lines in the top panel. Each symbol means different shock condition; n14v10 (circle), n13v20 (square), n12v35 (triangle), and n11v55 (inverse triangle), respectively. The isotopic fractionation should occur if the heating rate is lower than about $10^4 \, {\rm K/hr}$, which is indicated by a dashed line in the bottom panel. \label{fig:Tpre}}
\end{figure}

\begin{figure}
\epsscale{.90}
\plotone{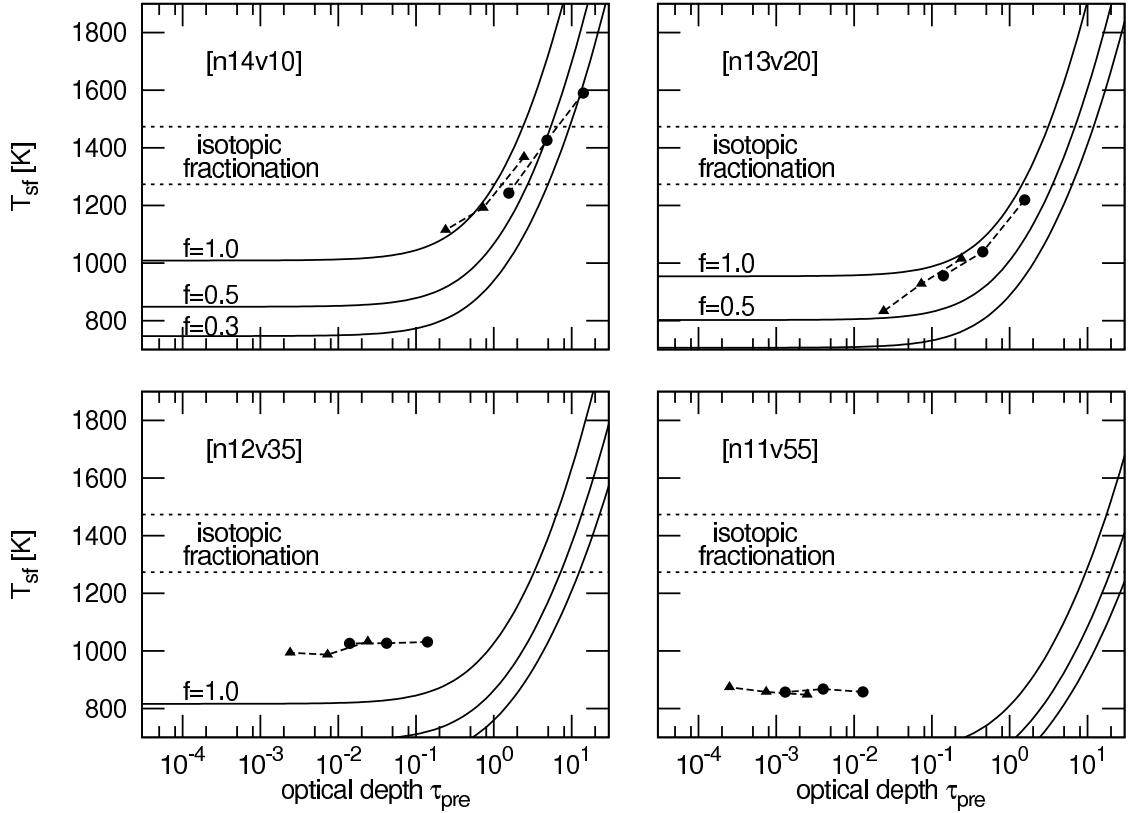}
\caption{Comparison of the pre-shock dust temperature estimated by using the radiative diffusion approximation (solid curve) with numerically calculated results (filled symbols) for various shock conditions: n14v10 (top left), n13v20 (top right), n12v35 (bottom left), and n11v55 (bottom right), respectively. Triangles indicate the results for the log-normal dust size distribution and circles are for power-law one. The parameter $f$ is the fraction of the net flux of the radiation returning from the shock front to the gas energy flux flowing into the shock front. \label{fig:Tpre2}}
\end{figure}

\begin{figure}
\epsscale{.90}
\plotone{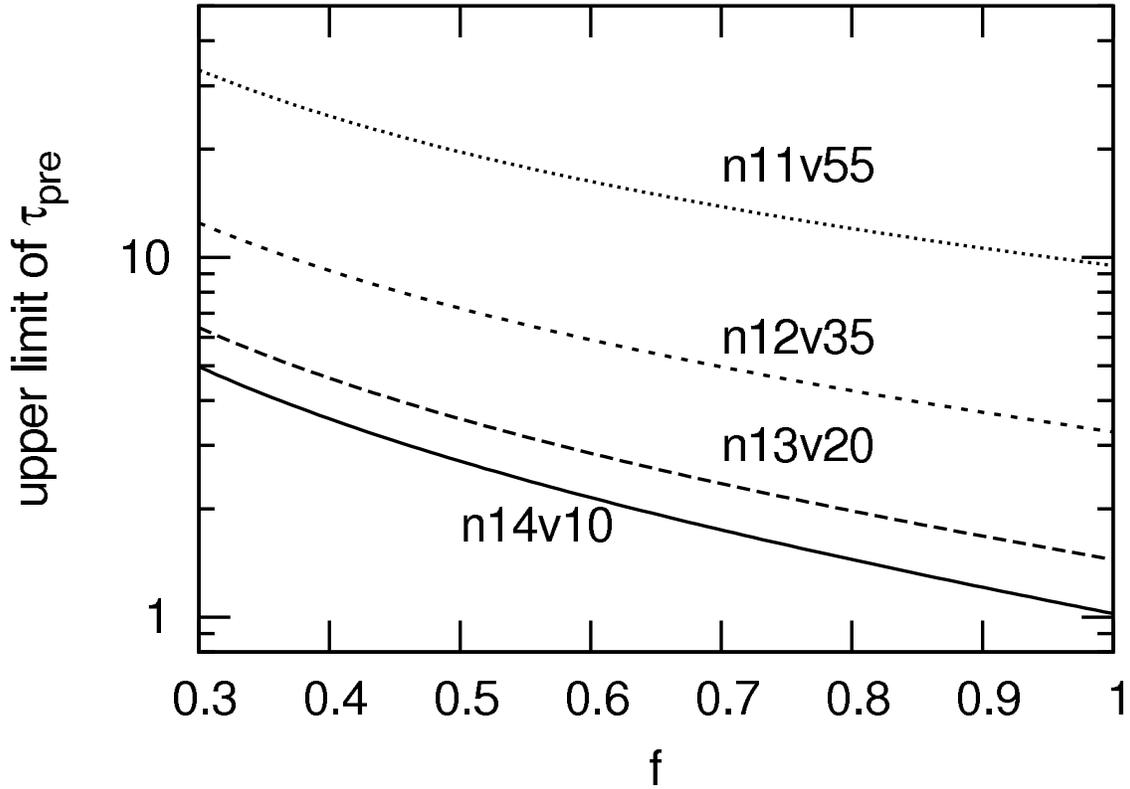}
\caption{Upper limit of the optical depth of the pre-shock region above which the isotopic fractionation should occur is plotted as a function of $f$, which is the fraction of the radiation flux returning from the shock front to the gas energy flux entering into the shock front. Labels indicate the shock condition of n14v10, n13v20, n12v35, and n11v55, respectively. \label{fig:tau_cr}}
\end{figure}

\begin{figure}
\epsscale{.70}
\plotone{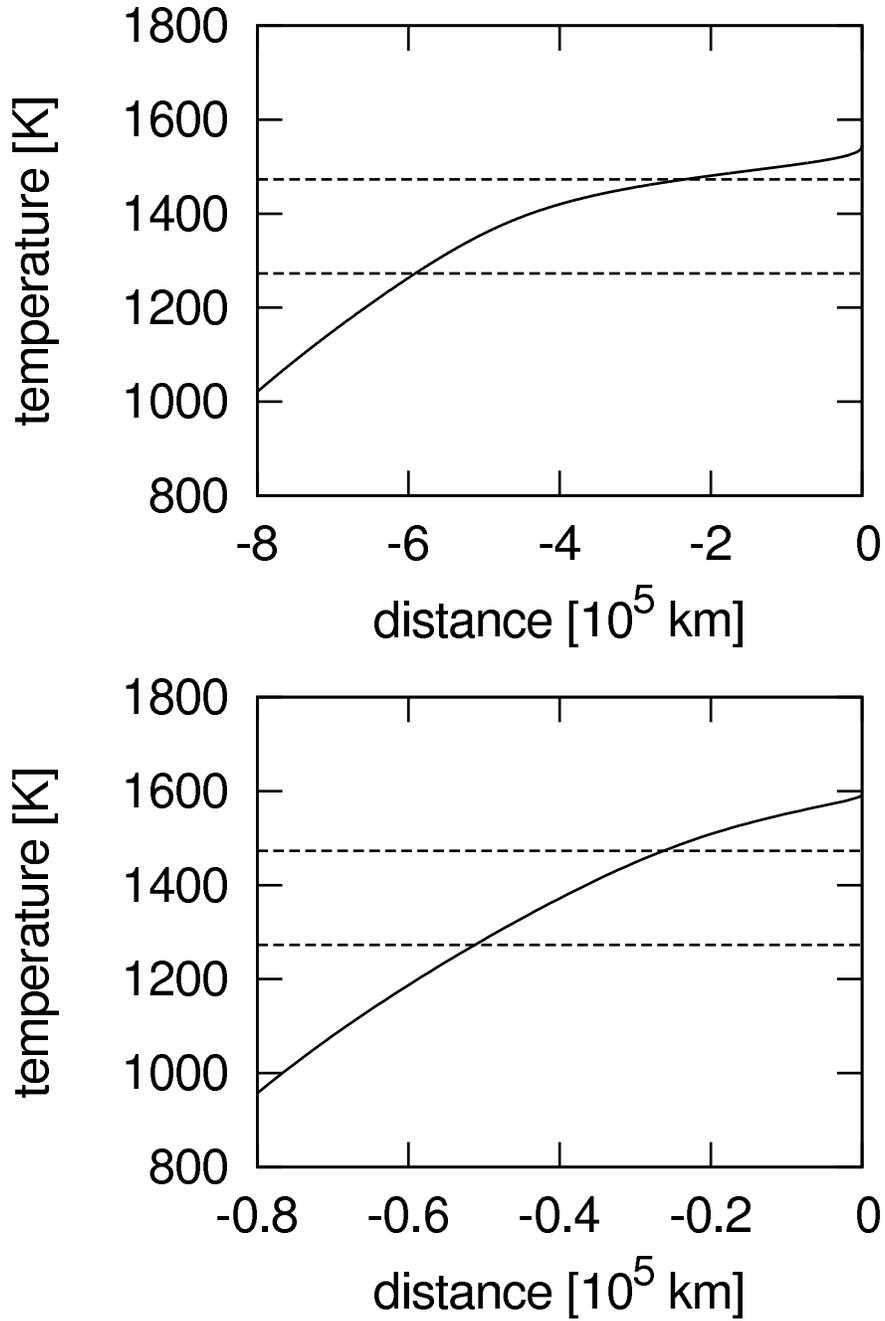}
\caption{The dust thermal histories in the pre-shock region for different spatial scale: $x_{\rm m} = 10^6 \, {\rm km}$ and $C_{\rm d} = 0.01$ (top), and $x_{\rm m} = 10^5 \, {\rm km}$ and $C_{\rm d} = 0.10$ (bottom). The shock condition is n14v10 and the dust size distribution is power-law. \label{fig:large_scale}}
\end{figure}

\begin{figure}
\epsscale{.80}
\plotone{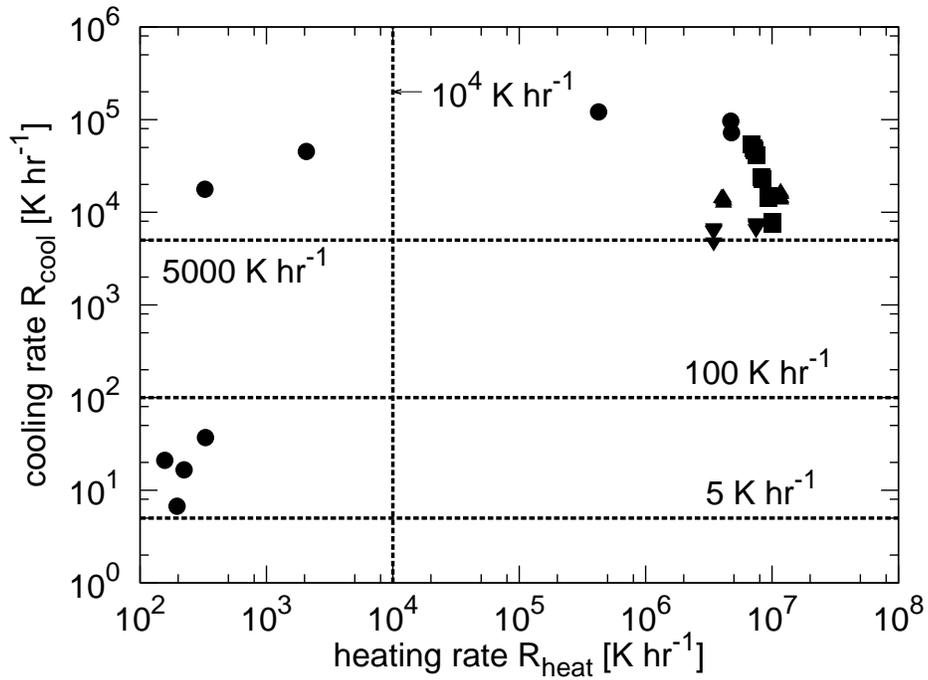}
\caption{Heating rates and cooling rates of chondrules for $a_0 = 100 \, {\rm \mu m}$. Each symbol means different gas number density of the pre-shock region; n14 (circle), n13 (square), n12 (triangle), and n11 (inverse triangle), respectively. \label{fig:heat_cool_rate_100}}
\end{figure}

\begin{figure}
\epsscale{.80}
\plotone{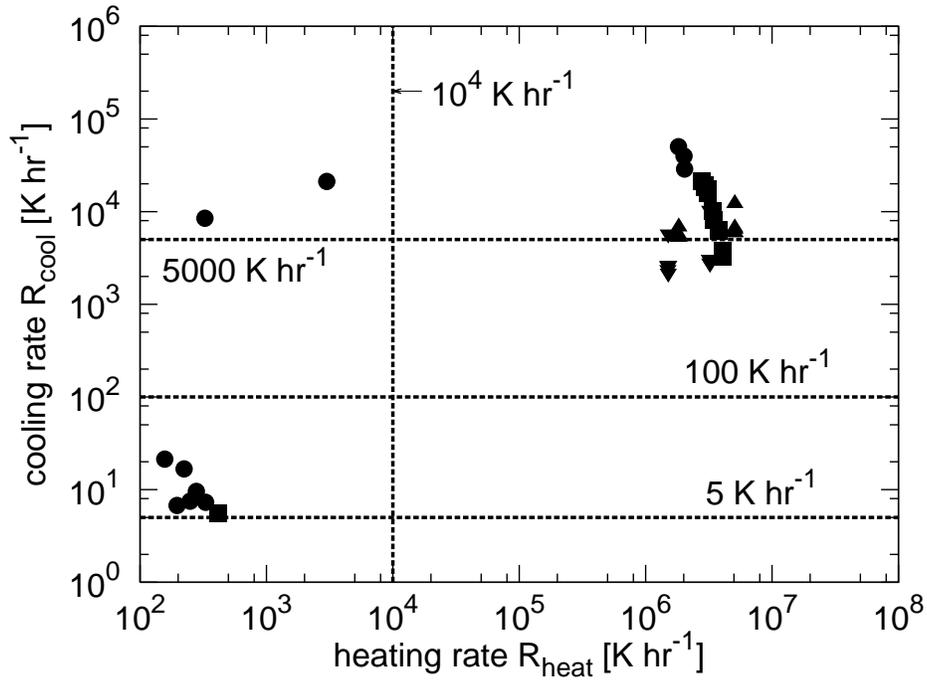}
\caption{Same as Fig. \ref{fig:heat_cool_rate_100} except for $a_0 = 251 \, {\rm \mu m}$. \label{fig:heat_cool_rate_251}}
\end{figure}

\begin{figure}
\epsscale{.80}
\plotone{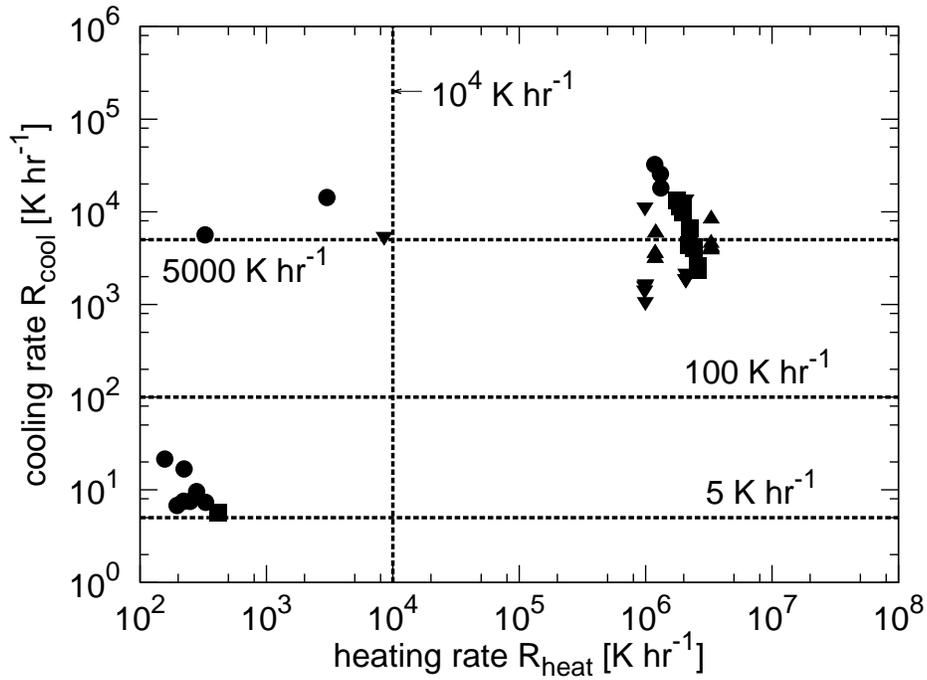}
\caption{Same as Fig. \ref{fig:heat_cool_rate_100} except for $a_0 = 398 \, {\rm \mu m}$. \label{fig:heat_cool_rate_398}}
\end{figure}

\begin{figure}
\epsscale{.80}
\plotone{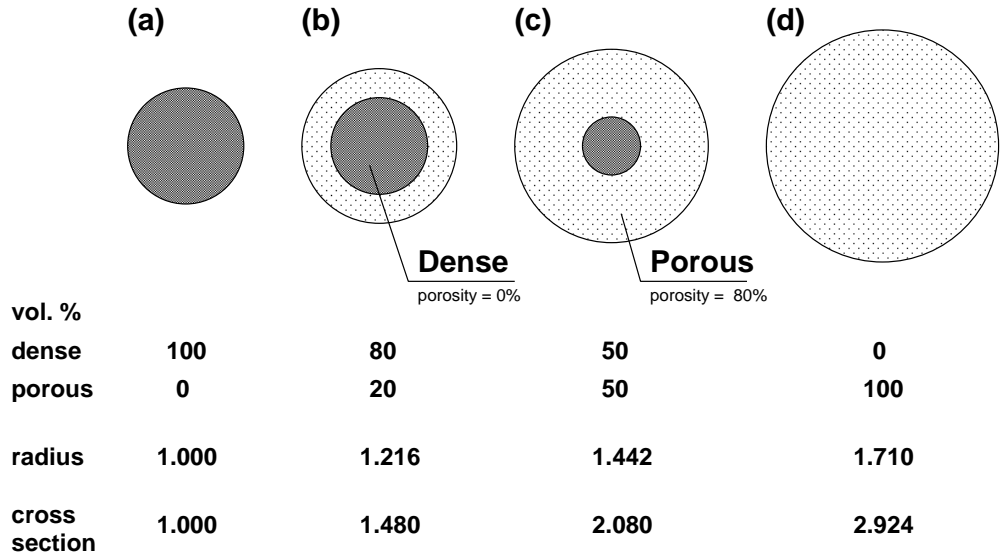}
\caption{Schematic pictures of precursor dust particles in various structures (fraction of the dense/porous region). The total masses of those particles are assumed to be the same. \label{fig:porosity}}
\end{figure}

\begin{figure}
\epsscale{1.}
\plotone{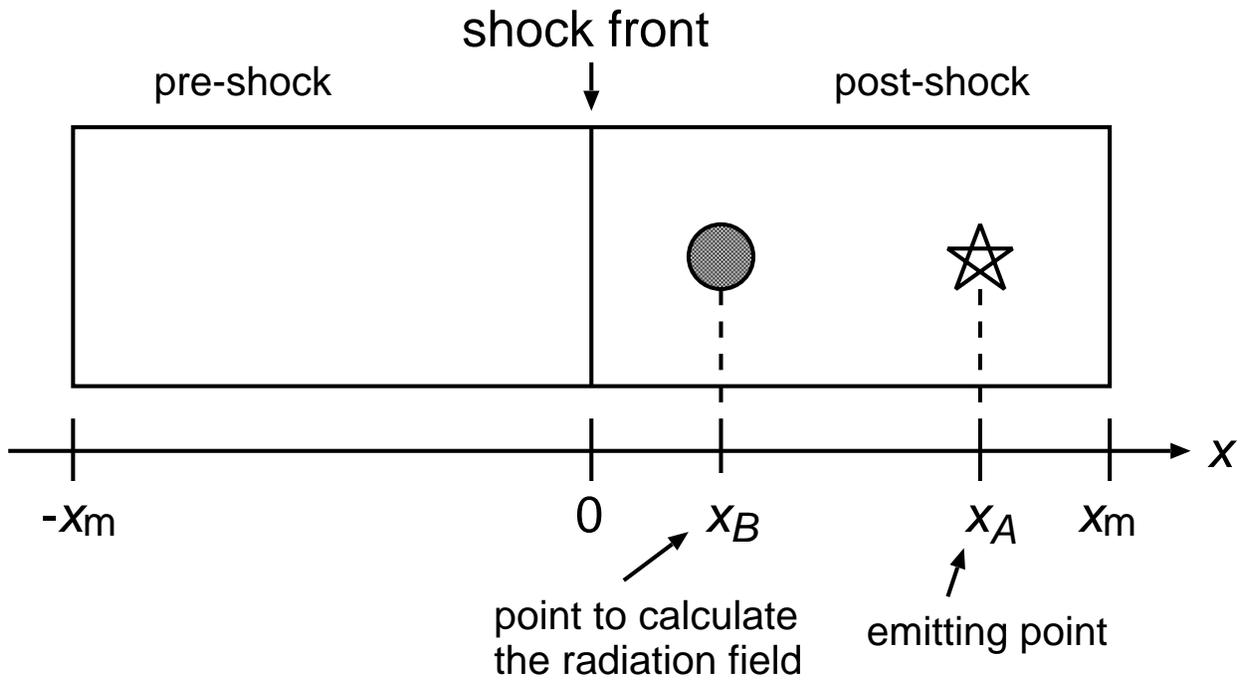}
\caption{Emitting point $x_A$ and a point at which we calculate the radiation field $x_B$. \label{fig:escape_prob}}
\end{figure}

Regarding the radiative flux in the post-shock region, there is another region. In the middle panel of Fig. \ref{fig:heatcoolrate_gas}, we found that the gas temperature decreases near the right boundary of the computational domain ($x \simeq x_{\rm m}$). From the bottom panel of Fig. \ref{fig:heatcoolrate_gas}, we also found that the cooling source is the dust particles, namely, the gas thermal energy is converted into the dust thermal energy by the thermal collision between the gas molecules and the dust surface. Since the dust particles are radiative equilibrium far behind the shock, the thermal energy transfered from the ambient gas radiates as the thermal radiation. Since the thermal emission works as the radiation source, the radiative flux is not constant at this region.




\clearpage


\begin{table}[]
\caption{Notation of the shock condition. The pre-shock gas number density is $n_0$ and the shock velocity is $v_{\rm s}$.}
\begin{center}
\begin{tabular}{|c||c|c|}
\hline
notation & $n_0$ [cm$^{-3}$] & $v_{\rm s}$ [km s$^{-1}$] \\
\hline
n14v08 & $10^{14}$ & 8 \\
n14v10 & $10^{14}$ & 10 \\
n14v13 & $10^{14}$ & 13 \\
\hline
n13v16 & $10^{13}$ & 16 \\
n13v20 & $10^{13}$ & 20 \\
n13v25 & $10^{13}$ & 25 \\
\hline
n12v30 & $10^{12}$ & 30 \\
n12v35 & $10^{12}$ & 35 \\
n12v40 & $10^{12}$ & 40 \\
\hline
n11v50 & $10^{11}$ & 50 \\
n11v55 & $10^{11}$ & 55 \\
n11v60 & $10^{11}$ & 60 \\
\hline
\end{tabular}
\end{center}
\label{table:shock_condition}
\end{table}

\begin{table}
\caption{Dust models we adopted in this study are listed. The charactor ``P" in the dust model means power-law size distribution and ``L" means lognormal one. The number ``10", ``03", and ``01" stand for the dust-to-gas mass ratio $C_{\rm d} = 0.10$, $0.03$, and $0.01$, respectively. $C_{\rm d}$.}
\begin{center}
\begin{tabular}{|c||c|c|}
\hline
notation & size dist. & $C_{\rm d}$ \\
\hline
P10 & power-law & 0.10 \\
P03 & power-law & 0.03 \\
P01 & power-law & 0.01 \\
\hline
L10 & lognormal & 0.10 \\
L03 & lognormal & 0.03 \\
L01 & lognormal & 0.01 \\
\hline
\end{tabular}
\end{center}
\label{table:dust_model}
\end{table}

\begin{table}
\caption{Calculation results of dust thermal histories of a dust particle of the initial radius $a_0 = 100 \, {\rm \mu m}$ are summarized; the optical depth of the pre-shock region $\tau_{\rm pre}$, the dust temperature just in front of the shock front $T_{\rm sf}$, the peak dust temperature $T_{\rm peak}$, the heating rate $R_{\rm heat}$, the cooling rate $R_{\rm cool}$, and the final dust radius $a_{\rm fin}$. $f$ indicates a ratio of the net flux of the radiation passing though the shock front to the gas energy flux lowing into the shock front. $a(b)=a\times10^b$. \label{table:summary_a100}}
\begin{center}
\small
\begin{tabular}{ccccccccc}
\hline
shock & dust & $\tau_{\rm pre}$ & $T_{\rm sf} \, {\rm [K]}$ & $T_{\rm peak} \, {\rm [K]}$ & $R_{\rm heat} \, {\rm [K/hr]}$ & $R_{\rm cool} \, {\rm [K/hr]}$ & $a_{\rm fin} \, {\rm [\mu m]}$ & $f$ \\
\hline \hline
n14v08 & P10 & 15.7 & 1469 & 1576 & 157 & 21.0 & 89 & 0.459 \\
 & P03 & 4.58 & 1241 & 1429 & --- & --- & 100 & 0.526 \\
 & P01 & 1.44 & 1059 & 1329 & --- & --- & 100 & 0.615 \\
 & L10 & 2.45 & 1153 & 1379 & --- & --- & 100 & 0.616 \\
 & L03 & 0.73 & 1023 & 1312 & --- & --- & 100 & 0.697 \\
 & L01 & 0.24 & 972 & 1289 & --- & --- & 100 & 0.711 \\
 \hline
n14v10 & P10 & 14.1 & 1590 & --- & --- & --- & 0 & 0.395 \\
 & P03 & 4.82 & 1426 & 1719 & 223 & 16.6 & 83 & 0.547 \\
 & P01 & 1.56 & 1244 & 1644 & 4.78(6) & 7.23(4) & 99 & 0.664 \\
 & L10 & 2.45 & 1369 & 1709 & 326 & 1.77(4) & 95 & 0.679 \\
 & L03 & 0.73 & 1192 & 1636 & 4.72(6) & 9.62(4) & 100 & 0.730 \\
 & L01 & 0.24 & 1116 & 1610 & 4.24(6) & 1.21(5) & 100 & 0.735 \\
\hline
n14v13 & P10 & 10.6 & 1709 & --- & --- & --- & 0 & 0.203 \\
 & P03 & 4.64 & 1551 & --- & --- & --- & 0 & 0.334 \\
 & P01 & 1.62 & 1389 & 2079 & 329 & 7.37 & 61 & 0.444 \\
 & L10 & 2.44 & 1607 & --- & --- & --- & 0 & 0.593 \\
 & L03 & 0.73 & 1419 & 2101 & 196 & 6.72 & 18 & 0.718 \\
 & L01 & 0.24 & 1305 & 2071 & 2.07(3) & 4.53(4) & 84 & 0.709 \\
\hline \hline
n13v16 & P10 & 1.44 & 1024 & 1526 & 1.98(6) & --- & 100 & 0.761 \\
 & P03 & 0.42 & 915 & 1500 & 1.50(6) & --- & 100 & 0.842 \\
 & P01 & 0.14 & 877 & 1492 & 1.34(6) & --- & 100 & 0.865 \\
 & L10 & 0.24 & 945 & 1508 & 1.67(6) & --- & 100 & 0.898 \\
 & L03 & 0.073 & 929 & 1504 & 1.59(6) & --- & 100 & 0.887 \\
 & L01 & 0.024 & 941 & 1507 & 1.65(6) & --- & 100 & 0.846 \\
\hline
n13v20 & P10 & 1.56 & 1219 & 1790 & 8.46(6) & 2.28(4) & 98 & 0.769 \\
 & P03 & 0.45 & 1039 & 1754 & 7.59(6) & 4.07(4) & 100 & 0.789 \\
 & P01 & 0.14 & 956 & 1741 & 7.28(6) & 4.71(4) & 100 & 0.788 \\
 & L10 & 0.24 & 1015 & 1753 & 7.56(6) & 4.37(4) & 100 & 0.848 \\
 & L03 & 0.073 & 928 & 1738 & 7.19(6) & 4.93(4) & 100 & 0.753 \\
 & L01 & 0.024 & 833 & 1725 & 6.88(6) & 5.40(4) & 100 & 0.487 \\
\hline
n13v25 & P10 & 1.67 & 1439 & 1951 & 414 & 5.50 & 55 & 0.680 \\
 & P03 & 0.48 & 1272 & 1891 & 1.01(7) & 7.56(3) & 93 & 0.758 \\
 & P01 & 0.16 & 1184 & 1866 & 9.32(6) & 1.51(4) & 99 & 0.735 \\
 & L10 & 0.24 & 1270 & 1892 & 1.01(7) & 7.96(3) & 94 & 0.885 \\
 & L03 & 0.073 & 1189 & 1869 & 9.39(6) & 1.44(4) & 99 & 0.768 \\
 & L01 & 0.024 & 1028 & 1835 & 8.29(6) & 2.40(4) & 100 & 0.249 \\
\hline
\end{tabular}
\end{center}
\label{default}
\end{table}

\addtocounter{table}{-1}
\begin{table}
\caption{Continue.}
\begin{center}
\small
\begin{tabular}{ccccccccc}
\hline
shock & dust & $\tau_{\rm pre}$ & $T_{\rm sf} \, {\rm [K]}$ & $T_{\rm peak} \, {\rm [K]}$ & $R_{\rm heat} \, {\rm [K/hr]}$ & $R_{\rm cool} \, {\rm [K/hr]}$ & $a_{\rm fin} \, {\rm [\mu m]}$ & $f$ \\
\hline \hline
n12v30 & P10 & 0.13 & 880 & 1361 & --- & --- & 100 & 0.919 \\
 & P03 & 0.040 & 871 & 1358 & --- & --- & 100 & 0.906 \\
 & P01 & 0.013 & 856 & 1355 & --- & --- & 100 & 0.725 \\
 & L10 & 0.025 & 844 & 1355 & --- & --- & 100 & 0.604 \\
 & L03 & 0.0074 & 808 & 1347 & --- & --- & 100 & 0.271 \\
 & L01 & 0.0025 & 816 & 1348 & --- & --- & 100 & 0.239 \\
\hline
n12v35 & P10 & 0.14 & 1054 & 1658 & 4.13(6) & 1.25(4) & 100 & 0.927 \\
 & P03 & 0.042 & 1050 & 1656 & 4.11(6) & 1.38(4) & 100 & 0.926 \\
 & P01 & 0.014 & 1049 & 1656 & 4.11(6) & 1.43(4) & 100 & 0.827 \\
 & L10 & 0.024 & 1056 & 1660 & 4.15(6) & 1.32(4) & 100 & 0.875 \\
 & L03 & 0.0073 & 1015 & 1650 & 3.99(6) & 1.41(4) & 100 & 0.391 \\
 & L01 & 0.0024 & 1022 & 1651 & 4.02(6) & 1.40(4) & 100 & 0.329 \\
\hline
n12v40 & P10 & 0.15 & 1219 & 1958 & 1.17(7) & 1.64(4) & 100 & 0.934 \\
 & P03 & 0.045 & 1217 & 1957 & 1.17(7) & 1.57(4) & 100 & 0.938 \\
 & P01 & 0.015 & 1220 & 1958 & 1.17(7) & 1.55(4) & 100 & 0.869 \\
 & L10 & 0.024 & 1228 & 1960 & 1.17(7) & 1.35(4) & 100 & 0.924 \\
 & L03 & 0.0073 & 1207 & 1955 & 1.17(7) & 1.41(4) & 100 & 0.575 \\
 & L01 & 0.0024 & 1203 & 1954 & 1.17(7) & 1.42(4) & 100 & 0.396 \\
\hline \hline
n11v50 & P10 & 0.013 & 808 & 1443 & --- & --- & 100 & 0.940 \\
 & P03 & 0.0039 & 813 & 1443 & --- & --- & 100 & 0.808 \\
 & P01 & 0.0013 & 803 & 1441 & --- & --- & 100 & 0.469 \\
 & L10 & 0.0025 & 793 & 1441 & --- & --- & 100 & 0.428 \\
 & L03 & 0.00074 & 805 & 1442 & --- & --- & 100 & 0.406 \\
 & L01 & 0.00025 & 819 & 1445 & --- & --- & 100 & 0.402 \\
\hline
n11v55 & P10 & 0.013 & 879 & 1622 & 3.45(6) & 4.81(3) & 100 & 0.950 \\
 & P03 & 0.0040 & 887 & 1622 & 3.47(6) & 6.43(3) & 100 & 0.857 \\
 & P01 & 0.0013 & 878 & 1620 & 3.45(6) & 6.84(3) & 100 & 0.506 \\
 & L10 & 0.0025 & 870 & 1620 & 3.43(6) & 6.38(3) & 100 & 0.473 \\
 & L03 & 0.00074 & 879 & 1621 & 3.47(6) & 6.63(3) & 100 & 0.426 \\
 & L01 & 0.00025 & 893 & 1623 & 3.45(6) & 6.60(3) & 100 & 0.420 \\
\hline
n11v60 & P10 & 0.013 & 948 & 1805 & 7.47(6) & 6.67(3) & 100 & 0.956 \\
 & P03 & 0.0041 & 958 & 1805 & 7.50(6) & 7.46(3) & 100 & 0.874 \\
 & P01 & 0.0014 & 949 & 1803 & 7.48(6) & 7.74(3) & 100 & 0.531 \\
 & L10 & 0.0024 & 965 & 1806 & 7.52(6) & 7.09(3) & 100 & 0.744 \\
 & L03 & 0.00073 & 951 & 1804 & 7.50(6) & 7.31(3) & 100 & 0.442 \\
 & L01 & 0.00024 & 966 & 1806 & 7.52(6) & 7.26(3) & 100 & 0.434 \\
\hline
\end{tabular}
\end{center}
\label{default}
\end{table}

\begin{table}
\caption{Same as Table \ref{table:summary_a100} except for the initial dust radius $a_0 = 251 \, {\rm \mu m}$. \label{table:summary_a251}}
\begin{center}
\small
\begin{tabular}{ccccccccc}
\hline
shock & dust & $\tau_{\rm pre}$ & $T_{\rm sf} \, {\rm [K]}$ & $T_{\rm peak} \, {\rm [K]}$ & $R_{\rm heat} \, {\rm [K/hr]}$ & $R_{\rm cool} \, {\rm [K/hr]}$ & $a_{\rm fin} \, {\rm [\mu m]}$ & $f$ \\
\hline \hline
n14v08 & P10 & 15.7 & 1469 & 1580 & 157 & 21.3 & 241 & 0.459 \\
 & P03 & 4.58 & 1241 & 1436 & --- & --- & 251 & 0.526 \\
 & P01 & 1.44 & 1058 & 1338 & --- & --- & 251 & 0.615 \\
 & L10 & 2.45 & 1152 & 1387 & --- & --- & 251 & 0.616 \\
 & L03 & 0.73 & 1022 & 1321 & --- & --- & 251 & 0.697 \\
 & L01 & 0.24 & 971 & 1298 & --- & --- & 251 & 0.711 \\
\hline
n14v10 & P10 & 14.1 & 1590 & 1821 & 279 & 9.57 & 145 & 0.395 \\
 & P03 & 4.82 & 1426 & 1728 & 223 & 16.7 & 234 & 0.547 \\
 & P01 & 1.56& 1243 & 1655 & 2.03(6) & 2.88(4) & 250 & 0.664 \\
 & L10 & 2.45 & 1368 & 1720 & 326 & 8.49(3) & 247 & 0.679 \\
 & L03 & 0.73 & 1191 & 1647 & 2.01(6) & 3.98(4) & 251 & 0.730 \\
 & L01 & 0.24 & 1115 & 1622 & 1.82(6) & 5.03(4) & 251 & 0.735 \\
\hline
n14v13 & P10 & 10.6 & 1709 & --- & --- & --- & 0 & 0.203 \\
 & P03 & 4.64 & 1551 & 2122 & 249 & 7.49 & 70 & 0.334 \\
 & P01 & 1.62& 1389 & 2086 & 330 & 7.30 & 211 & 0.444 \\
 & L10 & 2.44 & 1606 & --- & --- & --- & 0 & 0.593 \\
 & L03 & 0.73 & 1418 & 2107 & 196 & 6.74 & 167 & 0.718 \\
 & L01 & 0.24 & 1304 & 2077 & 3.00(3) & 2.12(4) & 234 & 0.709 \\
\hline \hline
n13v16 & P10 & 1.44 & 1024 & 1539 & 9.20(5) & --- & 251 & 0.761 \\
 & P03 & 0.42 & 915 & 1512 & 7.30(5) & --- & 251 & 0.842 \\
 & P01 & 0.14 & 877 & 1504 & 6.70(5) & --- & 251 & 0.865 \\
 & L10 & 0.24 & 944 & 1520 & 7.92(5) & --- & 251 & 0.898 \\
 & L03 & 0.073 & 928 & 1516 & 7.61(5) & --- & 251 & 0.887 \\
 & L01 & 0.024 & 940 & 1518 & 7.83(5) & --- & 251 & 0.846 \\
\hline
n13v20 & P10 & 1.56 & 1219 & 1798 & 3.44(6) & 8.18(3) & 249 & 0.769 \\
 & P03 & 0.45 & 1039 & 1761 & 3.09(6) & 1.58(4) & 251 & 0.789 \\
 & P01 & 0.14 & 956 & 1748 & 2.96(6) & 1.84(4) & 251 & 0.788 \\
 & L10 & 0.24 & 1015 & 1760 & 3.09(6) & 1.74(4) & 251 & 0.848 \\
 & L03 & 0.073 & 928 & 1744 & 2.93(6) & 1.95(4) & 251 & 0.753 \\
 & L01 & 0.024 & 833 & 1731 & 2.79(6) & 2.13(4) & 251 & 0.487 \\
\hline
n13v25 & P10 & 1.67 & 1439 & 1981 & 414 & 5.61 & 206 & 0.680 \\
 & P03 & 0.48 & 1271 & 1923 & 4.05(6) & 3.26(3) & 244 & 0.758 \\
 & P01 & 0.16 & 1183 & 1899 & 3.77(6) & 6.27(3) & 250 & 0.735 \\
 & L10 & 0.24 & 1269 & 1924 & 4.07(6) & 3.76(3) & 245 & 0.885 \\
 & L03 & 0.073 & 1187 & 1902 & 3.80(6) & 6.23(3) & 250 & 0.768 \\
 & L01 & 0.024 & 1024 & 1870 & 3.42(6) & 1.01(4) & 251 & 0.249 \\
\hline
\end{tabular}
\end{center}
\label{default}
\end{table}

\addtocounter{table}{-1}
\begin{table}
\caption{Continue.}
\begin{center}
\small
\begin{tabular}{ccccccccc}
\hline
shock & dust & $\tau_{\rm pre}$ & $T_{\rm sf} \, {\rm [K]}$ & $T_{\rm peak} \, {\rm [K]}$ & $R_{\rm heat} \, {\rm [K/hr]}$ & $R_{\rm cool} \, {\rm [K/hr]}$ & $a_{\rm fin} \, {\rm [\mu m]}$ & $f$ \\
\hline \hline
n12v30 & P10 & 0.13 & 871 & 1377 & --- & --- & 251 & 0.919 \\
 & P03 & 0.040 & 862 & 1372 & --- & --- & 251 & 0.906 \\
 & P01 & 0.013 & 848 & 1369 & --- & --- & 251 & 0.725 \\
 & L10 & 0.025 & 834 & 1370 & --- & --- & 251 & 0.604 \\
 & L03 & 0.0074 & 793 & 1362 & --- & --- & 251 & 0.271 \\
 & L01 & 0.0025 & 801 & 1363 & --- & --- & 251 & 0.239 \\
\hline
n12v35 & P10 & 0.14 & 1031 & 1670 & 1.83(6) & 6.82(3) & 251 & 0.927 \\
 & P03 & 0.042 & 1027 & 1667 & 1.82(6) & 5.26(3) & 251 & 0.926 \\
 & P01 & 0.014 & 1026 & 1667 & 1.82(6) & 5.67(3) & 251 & 0.827 \\
 & L10 & 0.024 & 1032 & 1673 & 1.84(6) & 5.32(3) & 251 & 0.875 \\
 & L03 & 0.0073 & 987 & 1663 & 1.77(6) & 5.75(3) & 251 & 0.391 \\
 & L01 & 0.0024 & 994 & 1665 & 1.78(6) & 5.71(3) & 251 & 0.329 \\
\hline
n12v40 & P10 & 0.15 & 1186 & 1969 & 5.07(6) & 1.22(4) & 251 & 0.934 \\
 & P03 & 0.045 & 1183 & 1968 & 5.07(6) & 6.65(3) & 251 & 0.938 \\
 & P01 & 0.015 & 1187 & 1968 & 5.07(6) & 6.42(3) & 251 & 0.869 \\
 & L10 & 0.024 & 1195 & 1973 & 5.08(6) & 5.91(3) & 251 & 0.924 \\
 & L03 & 0.0073 & 1171 & 1968 & 5.06(6) & 5.94(3) & 251 & 0.575 \\
 & L01 & 0.0024 & 1167 & 1967 & 5.05(6) & 5.92(3) & 251 & 0.396 \\
\hline \hline
n11v50 & P10 & 0.013 & 793 & 1492 & 2.82(3) & --- & 251 & 0.940 \\
 & P03 & 0.0039 & 799 & 1453 & --- & --- & 251 & 0.808 \\
 & P01 & 0.0013 & 785 & 1451 & --- & --- & 251 & 0.469 \\
 & L10 & 0.0025 & 774 & 1452 & --- & --- & 251 & 0.428 \\
 & L03 & 0.00074 & 787 & 1453 & --- & --- & 251 & 0.406 \\
 & L01 & 0.00025 & 805 & 1455 & --- & --- & 251 & 0.402 \\
\hline
n11v55 & P10 & 0.013 & 858 & 1692 & 1.51(6) & 5.76(3) & 251 & 0.950 \\
 & P03 & 0.0040 & 868 & 1633 & 1.51(6) & 2.18(3) & 251 & 0.857 \\
 & P01 & 0.0013 & 857 & 1631 & 1.50(6) & 2.63(3) & 251 & 0.506 \\
 & L10 & 0.0025 & 848 & 1632 & 1.50(6) & 2.38(3) & 251 & 0.473 \\
 & L03 & 0.00074 & 858 & 1633 & 1.50(6) & 2.61(3) & 251 & 0.426 \\
 & L01 & 0.00025 & 874 & 1635 & 1.52(6) & 2.65(3) & 251 & 0.420 \\
\hline
n11v60 & P10 & 0.013 & 924 & 1843 & 3.21(6) & 1.04(4) & 250 & 0.956 \\
 & P03 & 0.0041 & 935 & 1817 & 3.23(6) & 2.80(3) & 251 & 0.874 \\
 & P01 & 0.0014 & 926 & 1815 & 3.21(6) & 3.06(3) & 251 & 0.531 \\
 & L10 & 0.0024 & 926 & 1815 & 3.21(6) & 3.06(3) & 251 & 0.744 \\
 & L03 & 0.00073 & 935 & 1817 & 3.23(6) & 2.80(3) & 251 & 0.442 \\
 & L01 & 0.00024 & 926 & 1815 & 3.21(6) & 3.06(3) & 251 & 0.434 \\
\hline
\end{tabular}
\end{center}
\label{default}
\end{table}

\begin{table}
\caption{Same as Table \ref{table:summary_a100} except for the initial dust radius $a_0 = 398 \, {\rm \mu m}$. \label{table:summary_a398}}
\begin{center}
\small
\begin{tabular}{ccccccccc}
\hline
shock & dust & $\tau_{\rm pre}$ & $T_{\rm sf} \, {\rm [K]}$ & $T_{\rm peak} \, {\rm [K]}$ & $R_{\rm heat} \, {\rm [K/hr]}$ & $R_{\rm cool} \, {\rm [K/hr]}$ & $a_{\rm fin} \, {\rm [\mu m]}$ & $f$ \\
\hline \hline
n14v08 & P10 & 15.7 & 1469 & 1582 & 157 & 21.5 & 388 & 0.459 \\
 & P03 & 4.58 & 1241 & 1439 & --- & --- & 398 & 0.526 \\
 & P01 & 1.44 & 1058 & 1341 & --- & --- & 398 & 0.615 \\
 & L10 & 2.45 & 1152 & 1390 & --- & --- & 398 & 0.616 \\
 & L03 & 0.73 & 1021 & 1324 & --- & --- & 398 & 0.697 \\
 & L01 & 0.24 & 970 & 1301 & --- & --- & 398 & 0.711 \\
\hline
n14v10 & P10 & 14.1 & 1590 & 1823 & 280 & 9.57 & 292 & 0.395 \\
 & P03 & 4.82 & 1426 & 1731 & 223 & 16.8 & 381 & 0.547 \\
 & P01 & 1.56& 1243 & 1659 & 1.32(6) & 1.81(4) & 397 & 0.664 \\
 & L10 & 2.45 & 1368 & 1723 & 327 & 5.65(3) & 393 & 0.679 \\
 & L03 & 0.73 & 1191 & 1651 & 1.31(6) & 2.56(4) & 398 & 0.730 \\
 & L01 & 0.24 & 1115 & 1625 & 1.18(6) & 3.24(4) & 398 & 0.735 \\
\hline
n14v13 & P10 & 10.6 & 1709 & --- & --- & --- & 0 & 0.203 \\
 & P03 & 4.64 & 1551 & 2124 & 249 & 7.50 & 215 & 0.334 \\
 & P01 & 1.62& 1389 & 2087 & 330 & 7.31 & 357 & 0.444 \\
 & L10 & 2.44 & 1606 & 2166 & 221 & 7.53 & 63 & 0.593 \\
 & L03 & 0.73 & 1418 & 2107 & 196 & 6.76 & 313 & 0.718 \\
 & L01 & 0.24 & 1304 & 2077 & 3.01(3) & 1.43(4) & 380 & 0.709 \\
\hline \hline
n13v16 & P10 & 1.44 & 1024 & 1543 & 6.11(5) & --- & 398 & 0.761 \\
 & P03 & 0.42 & 915 & 1515 & 4.87(5) & --- & 398 & 0.842 \\
 & P01 & 0.14 & 877 & 1507 & 4.47(5) & --- & 398 & 0.865 \\
 & L10 & 0.24 & 943 & 1523 & 5.25(5) & --- & 398 & 0.898 \\
 & L03 & 0.073 & 927 & 1518 & 5.05(5) & --- & 398 & 0.887 \\
 & L01 & 0.024 & 940 & 1521 & 5.18(5) & --- & 398 & 0.846 \\
\hline
n13v20 & P10 & 1.56 & 1219 & 1802 & 2.18(6) & 4.41(3) & 396 & 0.769 \\
 & P03 & 0.45 & 1039 & 1764 & 1.96(6) & 9.95(3) & 398 & 0.789 \\
 & P01 & 0.14 & 956 & 1750 & 1.88(6) & 1.15(4) & 398 & 0.788 \\
 & L10 & 0.24 & 1015 & 1762 & 1.96(6) & 1.11(4) & 398 & 0.848 \\
 & L03 & 0.073 & 928 & 1746 & 1.85(6) & 1.23(4) & 398 & 0.753 \\
 & L01 & 0.024 & 833 & 1732 & 1.77(6) & 1.34(4) & 398 & 0.487 \\
\hline
n13v25 & P10 & 1.67 & 1438 & 1994 & 414 & 5.72 & 353 & 0.680 \\
 & P03 & 0.48 & 1270 & 1936 & 2.57(6) & 2.33(3) & 391 & 0.758 \\
 & P01 & 0.16 & 1182 & 1912 & 2.41(6) & 4.03(3) & 397 & 0.735 \\
 & L10 & 0.24 & 1268 & 1936 & 2.58(6) & 2.61(3) & 392 & 0.885 \\
 & L03 & 0.073 & 1186 & 1915 & 2.42(6) & 4.16(3) & 397 & 0.768 \\
 & L01 & 0.024 & 1021 & 1883 & 2.22(6) & 6.64(3) & 398 & 0.249 \\
\hline
\end{tabular}
\end{center}
\label{default}
\end{table}

\addtocounter{table}{-1}
\begin{table}
\caption{Continue.}
\begin{center}
\small
\begin{tabular}{ccccccccc}
\hline
shock & dust & $\tau_{\rm pre}$ & $T_{\rm sf} \, {\rm [K]}$ & $T_{\rm peak} \, {\rm [K]}$ & $R_{\rm heat} \, {\rm [K/hr]}$ & $R_{\rm cool} \, {\rm [K/hr]}$ & $a_{\rm fin} \, {\rm [\mu m]}$ & $f$ \\
\hline \hline
n12v30 & P10 & 0.13 & 865 & 1422 & --- & --- & 398 & 0.919 \\
 & P03 & 0.040 & 856 & 1377 & --- & --- & 398 & 0.906 \\
 & P01 & 0.013 & 842 & 1373 & --- & --- & 398 & 0.725 \\
 & L10 & 0.025 & 828 & 1375 & --- & --- & 398 & 0.604 \\
 & L03 & 0.0074 & 787 & 1367 & --- & --- & 398 & 0.271 \\
 & L01 & 0.0025 & 795 & 1369 & --- & --- & 398 & 0.239 \\
\hline
n12v35 & P10 & 0.14 & 1024 & 1689 & 1.20(6) & 5.90(3) & 398 & 0.927 \\
 & P03 & 0.042 & 1019 & 1670 & 1.19(6) & 3.14(3) & 398 & 0.926 \\
 & P01 & 0.014 & 1018 & 1669 & 1.19(6) & 3.55(3) & 398 & 0.827 \\
 & L10 & 0.024 & 1024 & 1689 & 1.20(6) & 5.90(3) & 398 & 0.875 \\
 & L03 & 0.0073 & 1019 & 1670 & 1.19(6) & 3.14(3) & 398 & 0.391 \\
 & L01 & 0.0024 & 1018 & 1669 & 1.19(6) & 3.55(3) & 398 & 0.329 \\
\hline
n12v40 & P10 & 0.15 & 1168 & 1972 & 3.30(6) & 8.34(3) & 397 & 0.934 \\
 & P03 & 0.045 & 1165 & 1970 & 3.29(6) & 4.57(3) & 398 & 0.938 \\
 & P01 & 0.015 & 1169 & 1970 & 3.30(6) & 4.15(3) & 398 & 0.869 \\
 & L10 & 0.024 & 1177 & 1977 & 3.31(6) & 4.04(3) & 398 & 0.924 \\
 & L03 & 0.0073 & 1152 & 1972 & 3.28(6) & 3.91(3) & 398 & 0.575 \\
 & L01 & 0.0024 & 1147 & 1971 & 3.28(6) & 3.87(3) & 398 & 0.396 \\
\hline \hline
n11v50 & P10 & 0.013 & 782 & 1627 & 8.54(3) & 5.46(3) & 398 & 0.940 \\
 & P03 & 0.0039 & 788 & 1456 & --- & --- & 398 & 0.808 \\
 & P01 & 0.0013 & 777 & 1454 & --- & --- & 398 & 0.469 \\
 & L10 & 0.0025 & 766 & 1456 & --- & --- & 398 & 0.428 \\
 & L03 & 0.00074 & 779 & 1457 & --- & --- & 398 & 0.406 \\
 & L01 & 0.00025 & 794 & 1459 & --- & --- & 398 & 0.402 \\
\hline
n11v55 & P10 & 0.013 & 850 & 1858 & 9.88(5) & 1.13(4) & 393 & 0.950 \\
 & P03 & 0.0040 & 859 & 1636 & 9.96(5) & 1.08(3) & 398 & 0.857 \\
 & P01 & 0.0013 & 849 & 1634 & 9.86(5) & 1.59(3) & 398 & 0.506 \\
 & L10 & 0.0025 & 840 & 1637 & 9.83(5) & 1.44(3) & 398 & 0.473 \\
 & L03 & 0.00074 & 850 & 1637 & 9.92(5) & 1.64(3) & 398 & 0.426 \\
 & L01 & 0.00025 & 865 & 1638 & 1.00(6) & 1.68(3) & 398 & 0.420 \\
\hline
n11v60 & P10 & 0.013 & 915 & 2019 & 2.08(6) & 1.38(4) & 380 & 0.956 \\
 & P03 & 0.0041 & 926 & 1821 & 2.09(6) & 2.18(3) & 398 & 0.874 \\
 & P01 & 0.0014 & 917 & 1819 & 2.08(6) & 1.90(3) & 398 & 0.531 \\
 & L10 & 0.0024 & 917 & 1819 & 2.08(6) & 1.90(3) & 398 & 0.744 \\
 & L03 & 0.00073 & 926 & 1821 & 2.09(6) & 2.18(3) & 398 & 0.442 \\
 & L01 & 0.00024 & 917 & 1819 & 2.08(6) & 1.90(3) & 398 & 0.434 \\
\hline
\end{tabular}
\end{center}
\label{default}
\end{table}

\clearpage




\clearpage

\end{document}